\documentclass[pr,onecolumn,10pt]{revtex4-2}
\usepackage[colorlinks,citecolor=blue,linkcolor=red]{hyperref}
\usepackage{amsmath,amssymb,graphicx,subfigure}
\usepackage{times}

\begin{document}

\title{Optical solitons and vortices in fractional media: A mini-review of
recent results}
\author{Boris A. Malomed$^{1,2}$}
\affiliation{$^{1}$Department of Physical Electronics, School of Electrical Engineering,
Faculty of Engineering, and Center for Light-Matter Interaction, Tel Aviv
University, P.O.B. 39040, Tel Aviv, Israel\\
$^2$Instituto de Alta Investigaci\'{o}n, Universidad de Tarapac\'{a},
Casilla 7D, Arica, Chile}

\begin{abstract}
The article produces a brief review of some recent results which predict
stable propagation of solitons and solitary vortices in models based on the
nonlinear Schr\"{o}dinger equation (NLSE) including fractional one- or
two-dimensional diffraction and cubic or cubic-quintic nonlinear terms, as
well as linear potentials. The fractional diffraction is represented by
fractional-order spatial derivatives of the Riesz type, defined in terms of
the direct and inverse Fourier transform. In this form, it can be realized
by spatial-domain light propagation in optical setups with a specially
devised combination of mirrors, lenses, and phase masks. The results
presented in the article were chiefly obtained in a numerical form. Some
analytical findings are included too -- in particular, for fast moving
solitons, and results produced by the variational approximation. Also
briefly considered are dissipative solitons which are governed by the
fractional complex Ginzburg-Landau equation.
\end{abstract}

\maketitle


\section{Introduction and the basic models}

Nonlinear Schr\"{o}dinger equations (NLSEs) give rise to soliton families in
a great number of realizations \cite{REV1}-\cite{REV11}, many of which
originate in optics. The introduction of fractional calculus in NLSEs has
drawn much interest since it was proposed -- originally, in the linear form
-- as the quantum-mechanical model, derived from the respective
Feynman-integral formulation, for particles moving by L\'{e}vy flights \cite%
{Lask1,Lask2}. Experimental implementation of fractional linear Schr\"{o}%
dinger equations has been reported in condensed matter \cite{EXP1,EXP2} and
photonics \cite{EXP3}, in the form of transverse dynamics in optical
cavities. Realization of the propagation dynamics of light beams governed by
this equation was proposed in Ref. \cite{PROP}, followed by its extension
for models including complex potentials subject to the condition of the
parity-time ($\mathcal{PT}$) symmetry \cite{PTS,ghosts}. Generally, optical
settings modeled by linear and nonlinear fractional Schr\"{o}dinger
equations may be considered as a specific form of artificial photonic media.

The modulational instability of continuous waves \cite{Conti} and many types
of optical solitons produced by fractional NLSEs \cite{soliton1}-\cite%
{soliton5} have been theoretically investigated, chiefly by means of
numerical methods. These are quasi-linear \textquotedblleft accessible
solitons" (actually, quasi-linear modes) \cite{Frac1,Frac2}, gap solitons
supported by spatially periodic (lattice) potentials \cite{Frac5a}-\cite%
{Frac5}, solitary vortices \cite{Frac6,Frac7}, multipole and multi-peak
solitons \cite{Frac8}-\cite{Frac11}, soliton clusters \cite{Frac12},
solitary states with spontaneously broken symmetry \cite%
{Frac15,Frac16,Frac17}, as well as solitons in optical couplers \cite%
{Frac18,Frac19}. Dissipative solitons in fractional complex Ginzburg-Landau
equation (CGLE) were studied too \cite{Frac14}.

The objective of this article is to present a short review of models based
on fractional NLSEs and some states produced by them. The review is not
drafted to be a comprehensive one, the bibliography not being comprehensive
either; rather, it selects several recent results which seem interesting in
terms of the general soliton dynamics.

The basic form of the one-dimensional (1D) fractional NLSE model, written
for amplitude $\Psi $ of the optical field, is
\begin{equation}
i\frac{\partial \Psi }{\partial z}=\frac{1}{2}\left( -\frac{\partial ^{2}}{%
\partial x^{2}}\right) ^{\alpha /2}\Psi +V(x)\Psi -g\left\vert \Psi
\right\vert ^{2}\Psi .  \label{NLFSE}
\end{equation}%
Here $z$ and $x$ are the scaled propagation distance and transverse
coordinate, $g>0$ ($g<0$) is the coefficient representing the self-focusing
(defocusing) cubic (Kerr) nonlinearity, and $V(x)$ is a trapping potential
which may be included in the model. The fractional-diffraction operator with
the \textit{L\'{e}vy index} $\alpha $ (in the original fractional Schr\"{o}%
dinger equation, it characterizes the hopping motion of the quantum particle
\cite{Lask1}) is defined, in the form of the \textit{Riesz derivative} \cite%
{Riesz}, by means of the juxtaposition of the direct and inverse Fourier
transform \cite{Lask1}-\cite{EXP3}, \cite{soliton2}:
\begin{equation}
\left( -\frac{\partial ^{2}}{\partial x^{2}}\right) ^{\alpha /2}\Psi =\frac{1%
}{2\pi }\int_{-\infty }^{+\infty }dp|p|^{\alpha }\int_{-\infty }^{+\infty
}d\xi e^{ip(x-\xi )}\Psi (\xi ),  \label{FracDefi}
\end{equation}%
In particular, the above-mentioned concatenation of mirrors, lenses, and
phase masks in optical cavities makes it possible to implement the effective
fractional diffraction approximated by operator (\ref{FracDefi}) \cite{EXP3}.

The L\'{e}vy index in Eq. (\ref{NLFSE}) may take values
\begin{equation}
1<\alpha \leq 2,  \label{12}
\end{equation}%
where $\alpha =2$ corresponds to the normal paraxial diffraction,
represented by operator $-\partial ^{2}/\partial x^{2}$. Equation (\ref%
{NLFSE}) with $g>0$ (self-focusing) gives rise to the \textit{critical
collapse} at $\alpha =1$ (and to the \textit{supercritical collapse} at $%
\alpha <1$), which destabilizes all possible soliton solutions \cite%
{Chen,Frac14}, therefore only values $\alpha >1$ are usually considered (the
linear fractional Schr\"{o}dinger equation with $\alpha =1$ admits
analytical solutions based on Airy functions \cite{EXP3}).

In the case of quadratic self-focusing nonlinearity, the critical collapse
occurs at $\alpha =1/2$, hence the L\'{e}vy index may take values $%
1/2<\alpha \leq 2$ in that case. The quadratic self-focusing term appears on
the right-hand side of Eq. (\ref{NLFSE}), in the form of $-\varepsilon |\psi
|\psi $ with $\varepsilon >0$, as the correction induced by quantum
fluctuations \cite{PA} to the 1D fractional Gross-Pitaevskii equation for
the gas of L\'{e}vy-hopping particles \cite{Liangwei}.

Besides the real waveguiding potential, $V(x)$, Eq. (\ref{NLFSE}) may
include a\ $\mathcal{PT}$-symmetric complex potential, with even real and
odd imaginary parts:%
\begin{equation}
V^{\ast }(x)=V(-x),  \label{PT}
\end{equation}%
where $\ast $ stands for the complex conjugate. As usual, in terms of
optical waveguides the odd imaginary part of potential (\ref{PT}) represents
spatially separated and mutually balanced gain and loss elements.

The two-dimensional (2D) version of the underlying equation (\ref{NLFSE}),
with two transverse coordinates, $x$ and $y$, is written as
\begin{equation}
i\frac{\partial \Psi }{\partial z}=\frac{1}{2}\left( -\frac{\partial ^{2}}{%
\partial x^{2}}-\frac{\partial ^{2}}{\partial y^{2}}\right) ^{\alpha /2}\Psi
+V(x,y)\Psi -g\left\vert \Psi \right\vert ^{2}\Psi .  \label{2+1D}
\end{equation}%
In this case, the fractional-diffraction operator is defined as%
\begin{equation}
\left( -\frac{\partial ^{2}}{\partial x^{2}}-\frac{\partial ^{2}}{\partial
y^{2}}\right) ^{\alpha /2}\Psi =\frac{1}{(2\pi )^{2}}\int \int dpdq\left(
p^{2}+q^{2}\right) ^{\alpha /2}\int \int d\xi d\eta e^{i\left[ p(x-\xi
)+iq(y-\eta )\right] }\Psi (\xi ,\eta ),  \label{2D operator}
\end{equation}%
cf. Eq. (\ref{FracDefi}).

The rest of the article is organized as follows. In Section II, basic
numerical and analytical results are presented for soliton families
generated by the 1D fractional NLSE (\ref{NLFSE}) in the free space (without
the trapping potential, $V=0$). Other essential numerical findings for 1D
solitons are presented in Section III. In particular, these are results
produced by trapping potentials, and dissipative solitons obtained in the
framework of the fractional complex Ginzburg-Landau equation (CGLE) with the
cubic-quintic nonlinearity. Selected results for 2D solitons, including ones
with embedded vorticity and necklace-shaped soliton clusters, are collected
in Section IV. To secure stability of the solitons, the 2D model includes
the quintic self-defocusing term, in addition to the cubic self-focusing
one; the stabilization of 2D solitons can also be provided by the parabolic
trapping potential, see Eqs. (\ref{Omega}) and (\ref{final}) below. The
paper is concluded by Section V.

\section{Basic soliton families in the one-dimensional fractional medium}

\subsection{Stationary states and analysis of their stability}

Steady-state solutions to Eq. (\ref{NLFSE}) with propagation constant $-\mu $
are looked for in the form of
\begin{equation}
\Psi (x,z)=U(x)\mathrm{exp}(-i\mu z),  \label{SSE}
\end{equation}%
with real function $U(x)$ satisfying the stationary equation,
\begin{equation}
\mu U=\frac{1}{2}\left( -\frac{\partial ^{2}}{\partial x^{2}}\right)
^{\alpha /2}U-gU^{3}  \label{NLFSES}
\end{equation}%
(here, the external potential is dropped, $V=0$). The stationary states are
characterized by their power (alias norm),
\begin{equation}
N=\int_{-\infty }^{+\infty }\left\vert U(x)\right\vert ^{2}dx.  \label{P}
\end{equation}

Stability of these states was investigated by taking the perturbed solution
as
\begin{equation}
\Psi =[U(x)+a(x)\mathrm{exp}(\lambda z)+b^{\ast }(x)\mathrm{exp}(\lambda
^{\ast }z)]\mathrm{exp}(-i\mu z),  \label{PERB}
\end{equation}%
where $a(x)$ and $b^{\ast }(x)$ are components of small complex
perturbations, and $\lambda $ is an instability growth rate (which may be
complex). Substituting this ansatz in Eq. (\ref{NLFSE}), one derives the
linearized equations for $a$ and $b$:
\begin{eqnarray}
i\lambda a &=&\frac{1}{2}\left( -\frac{\partial ^{2}}{\partial x^{2}}\right)
^{\alpha /2}a-\mu a+gU^{2}(2a+b),  \notag \\
i\lambda b &=&-\frac{1}{2}\left( -\frac{\partial ^{2}}{\partial x^{2}}%
\right) ^{\alpha /2}b+\mu b-gU^{2}(2b+a).  \label{LAS}
\end{eqnarray}%
The underlying stationary solution (\ref{SSE}) is stable provided that all
eigenvalues $\lambda $ produced by Eq. (\ref{LAS}) have zero real parts.

\subsection{The quasi-local approximation for modes carried by a rapidly
oscillating continuous wave}

To better understand the purport of the fractional-diffraction operator
defined by Eq. (\ref{FracDefi}), it is instructive to consider its action on
function $\Psi (x)$ built as a slowly varying envelope $\psi (x)$
multiplying a rapidly oscillating continuous-wave carrier,%
\begin{equation}
\Psi \left( x\right) =\psi (x)e^{iPx},  \label{psi}
\end{equation}%
with large wavenumber $P$. In particular, this ansatz may be used to
construct solutions for rapidly moving modes, see Eq. (\ref{Vgr}) below (in
fact, in the spatial domain the \textquotedblleft moving" modes are ones
tilted in the $\left( x,z\right) $ plane). The substitution of wave form (%
\ref{psi}) in the nonlocal expression on the right-hand side of Eq. (\ref%
{FracDefi}) readily leads to the following result, in the form of a
quasi-local expression expanded in powers of small parameter $1/P$:
\begin{equation}
\left( -\frac{\partial ^{2}}{\partial x^{2}}\right) ^{\alpha /2}\left( \psi
(x)e^{iPx}\right) =e^{iPx}|P|^{\alpha }\left[ \psi +\sum_{n=1}^{\infty
}(-i)^{n}\frac{\alpha \left( \alpha -1\right) ...\left( \alpha -n+1\right) }{%
n!P^{n}}\frac{\partial ^{n}\psi }{\partial x^{n}}\right] .  \label{quasi}
\end{equation}

The substitution of expansion (\ref{quasi}), truncated at some $n=n_{\max }$%
, in Eq. (\ref{NLFSE}) and cancellation of the common factor $e^{iPx}$ leads
to the familiar NLSE with higher-order dispersion (diffraction) terms, which
correspond to $n\geq 3$ in Eq. (\ref{quasi}). In particular, setting $\psi =%
\mathrm{const}$ demonstrates that the dispersion relation produced by the
linearized version of Eq. (\ref{NLFSES}) is
\begin{equation}
\mu =(1/2)|P|^{\alpha },  \label{mu}
\end{equation}%
in agreement with the analysis of Laskin \cite{Lask2}. Equation (\ref{mu})
implies that the semi-infinite bandgap, with $\mu <0$, may be populated by
solitons, in the framework of the nonlinear equation.

As concerns the term $\sim i\partial \psi /\partial x$, corresponding to $%
n=1 $, which appears after the substitution of expansion (\ref{quasi}) in
Eq. (\ref{NLFSE}), it suggests one to rewrite the NLSE in the reference
frame moving with the group velocity%
\begin{equation}
V_{\mathrm{gr}}=\left( \alpha /2\right) |P|^{\alpha -1}\mathrm{sgn}(P),
\label{Vgr}
\end{equation}%
i.e., to replace $x$ by
\begin{equation}
\tilde{x}\equiv x-V_{\mathrm{gr}}z  \label{tilde}
\end{equation}%
in the resulting equation (in fact, the group velocity is the tilt of
spatial solitons). Note that, for values of the L\'{e}vy index belonging to
interval (\ref{12}), the group velocity is large for large $P$, while the
effective second-order diffraction coefficient, as determined by the term
with $n=2$ in Eq. (\ref{quasi}), is small,%
\begin{equation}
D_{2}=(1/2)\alpha \left( \alpha -1\right) |P|^{-\left( 2-\alpha \right) }.
\label{D2}
\end{equation}%
The respectively transformed NLSE (\ref{NLFSE}) (without the external
potential, $V(x)=0$), truncated at $n_{\max }=2$, is%
\begin{equation}
i\frac{\partial \tilde{\psi}}{\partial z}=-\frac{1}{2}D_{2}\frac{\partial
^{2}\tilde{\psi}}{\partial \tilde{x}^{2}}-g\left\vert \tilde{\psi}%
\right\vert ^{2}\tilde{\psi},  \label{psitilde}
\end{equation}%
where $\tilde{\psi}\left( \tilde{x},z\right) =\exp \left( (i/2)|P|^{\alpha
}z\right) \cdot \psi \left( x,z\right) $. The obvious soliton solution of
Eq. (\ref{psitilde}) with the self-focusing nonlinearity, $g>0$, written in
terms of norm (\ref{P}), is%
\begin{equation}
\tilde{\psi}=\exp \left( i\frac{(gN)^{2}}{8D_{2}}z\right) \frac{N}{2}\sqrt{%
\frac{g}{D_{2}}}\mathrm{sech}\left( \frac{N}{2}\frac{g}{D_{2}}\tilde{x}%
\right) .  \label{soliton}
\end{equation}%
It follows from Eq. (\ref{D2}) that this soliton is a narrow one, with the
width estimated as $W\sim N^{-1}|P|^{-\left( 2-\alpha \right) }$. Further,
in this case the terms representing higher-order diffraction terms with $%
n\geq 3$, which originate from the expansion (\ref{quasi}), are relatively
small perturbations decaying $\sim |P|^{-n\left( \alpha -1\right) }$.

\subsection{The scaling relation and variational approximation (VA) for
soliton families}

Equation (\ref{NLFSES}) gives rise to an exact scaling relation between the
soliton's power and propagation constant:%
\begin{equation}
N(\mu ,g)=N_{0}(\alpha )g^{-1}\left( -\mu \right) ^{1-1/\alpha },
\label{cubic}
\end{equation}%
with a constant $N_{0}(\alpha )$ (its particular value is $N_{0}(\alpha =2)=2%
\sqrt{2}$, see Eq. (\ref{cubic-usual}) below). The fact that relation (\ref%
{cubic}) satisfies the commonly known \textit{Vakhitov-Kolokolov} (VK)\
criterion,
\begin{equation}
dN/d\mu <0  \label{VaKo}
\end{equation}
\cite{VK,Berge} at $\alpha >1$ suggests that the respective soliton family
may be stable, see further details in Fig. \ref{fig3} below. The case of $%
\alpha =1$, which corresponds to the degenerate form of relation (\ref{cubic}%
), with $N(\mu )\equiv \mathrm{const}$, implies the occurrence of the
above-mentioned critical collapse, which makes all solitons unstable (cf.
the commonly known cubic NLSE in the 2D space with the normal diffraction, $%
\alpha =2$, in which the family of \textit{Townes solitons}, destabilized by
the critical collapse, has a single value of the norm \cite{Berge}). In the
case of $\alpha <1$, the solitons generated by Eq. (\ref{NLFSES}) are made
strongly unstable by the presence of the supercritical collapse, similar to
solitons of the usual cubic NLSE in three dimensions \cite{Berge}.

Localized solutions of the fractional NLSE can be looked for in an
approximate analytical form by means of the variational approximation (VA)
\cite{Chen,Frac14,we}. To introduce it, note that Eq. (\ref{NLFSES}) for
real $U(x)$, with the fractional diffraction operator defined as per Eq. (%
\ref{FracDefi}), can be derived from the Lagrangian,%
\begin{gather}
L=-\frac{\mu }{2}\int_{-\infty }^{+\infty }dxU^{2}(x)+\frac{1}{8\pi }%
\int_{-\infty }^{+\infty }dp|p|^{\alpha }\int \int d\xi dxe^{ip(x-\xi
)}U(x)U(\xi )  \notag \\
-\frac{g}{4}\int_{-\infty }^{+\infty }dxU^{4}(x).  \label{Lagr}
\end{gather}%
The simplest form of the variational \textit{ansatz} approximating the
solitons sought for is based on the Gaussian (cf. Ref. \cite{Anderson}),%
\begin{equation}
\mathcal{U}(x)=A\exp \left( -\frac{x^{2}}{2W^{2}}\right) ,  \label{ans}
\end{equation}%
with real amplitude $A$, width $W$, and the power calculated as per Eq. (\ref%
{P}),%
\begin{equation}
\mathcal{N}=\sqrt{\pi }A^{2}W  \label{Nans}
\end{equation}%
(the calligraphic font denotes quantities pertaining to VA). The
substitution of the ansatz in Lagrangian (\ref{Lagr}) yields the
corresponding effective Lagrangian, which may be conveniently written with
squared amplitude $A^{2}$ replaced by the norm, according to Eq. (\ref{Nans}%
) \cite{Frac14}:
\begin{equation}
L_{\mathrm{eff}}=-\frac{\mu }{2}\mathcal{N}+\frac{\Gamma \left( \left(
1+\alpha \right) /2\right) }{4\sqrt{\pi }}\frac{\mathcal{N}}{W^{\alpha }}-%
\frac{g}{4\sqrt{2\pi }}\frac{\mathcal{N}^{2}}{W},  \label{Leff}
\end{equation}%
where $\Gamma $ is the Gamma-function. Then, values of $\mathcal{N}$ and $W$
are predicted by the Euler-Lagrange equations,%
\begin{equation}
\partial L_{\mathrm{eff}}/\partial \mathcal{N}=\partial L_{\mathrm{eff}%
}/\partial W=0.  \label{EL}
\end{equation}

\begin{figure}[tbp]
\begin{center}
\includegraphics[width=0.76\columnwidth]{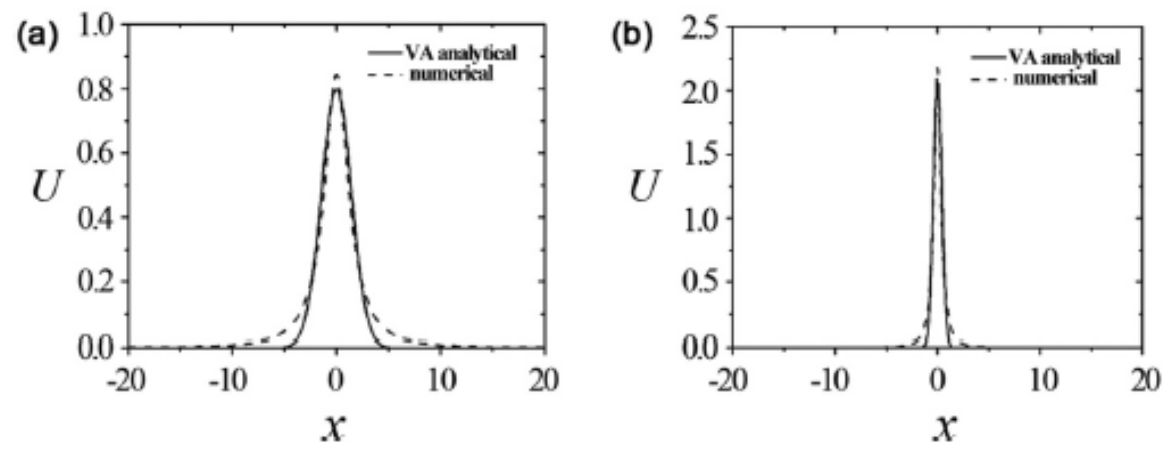}
\end{center}
\caption{Profiles of solitons with $\protect\mu =-0.3$ (a) and $-2.0$ (b),
predicted by the VA based on the Gaussian ansatz \ (\protect\ref{ans}) and
Euler-Lagrange equations (\protect\ref{EL}), and their counterparts produced
by the numerical solution of Eq. (\protect\ref{NLFSES}) with L\'{e}vy index $%
\protect\alpha =1.5$ and $g=1$. Reprinted with permission from Ref.
\protect\cite{Frac14}. Copyright 2020 Elsevier.}
\label{fig1}
\end{figure}

\begin{figure}[tbp]
\begin{center}
\includegraphics[width=0.42\columnwidth]{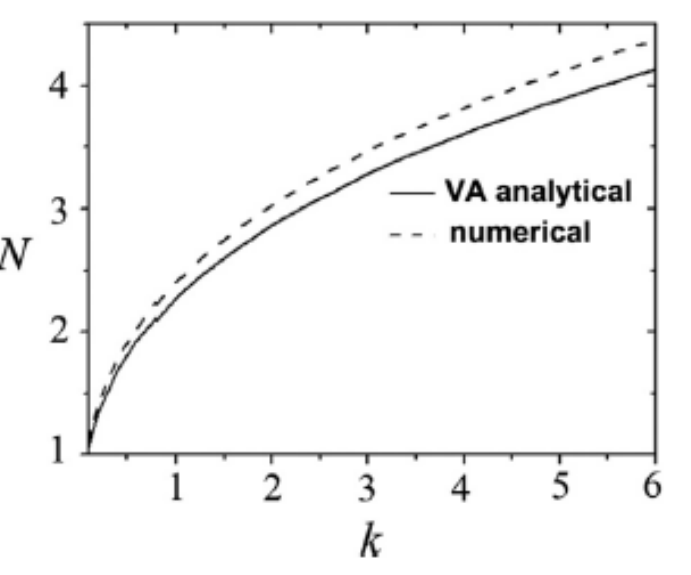}
\end{center}
\caption{Dependence of the soliton's norm $N$ on the propagation constant, $%
k\equiv -\protect\mu $, for the L\'{e}vy index $\protect\alpha =1.5$ and
self-focusing coefficient $g=1$, as predicted by the VA based on the
Gaussian ansatz \ (\protect\ref{ans}) and Euler-Lagrange equations (\protect
\ref{EL}). The corresponding dependence obtained from the numerical solution
of Eq. (\protect\ref{NLFSES}) is included too. Reprinted with permission
from Ref. \protect\cite{Frac14}. Copyright 2020 Elsevier.}
\label{fig2}
\end{figure}

Particular examples of the solitons shapes predicted by the VA, and their
comparison to the numerically found counterparts are shown in Fig. \ref{fig1}%
. At a fixed value of the L\'{e}vy index, soliton families are characterized
by dependences $N(\mu )$. An example of such a VA-predicted dependence and
its numerical counterpart are displayed in Fig. \ref{fig2} for $\alpha =1.5$%
, which demonstrates a sufficiently good accuracy of the VA.

An essential result of the VA is the prediction of the fixed norm for the
\textit{quasi-Townes} solitons, for which the critical collapse takes place
in the free space at $\alpha =1$ \cite{Frac14}:%
\begin{equation}
\left( \mathcal{N}_{\mathrm{Townes}}\right) _{\mathrm{VA}}=\sqrt{2}\approx
1.41.  \label{cubic-VA}
\end{equation}%
This result is similar to the well-known VA prediction for the norm of the
Townes solitons in the 2D NLSE with the cubic self-focusing \cite{Anderson},
\begin{equation}
\left( \mathcal{N}_{\mathrm{Townes}}^{\mathrm{(2D)}}\right) _{\mathrm{VA}%
}=2\pi  \label{VA-TS}
\end{equation}
\cite{Anderson}, while the respective numerical value is
\begin{equation}
N_{\mathrm{Townes}}^{\mathrm{(2D)}}\approx 5.85  \label{Townes}
\end{equation}%
\cite{Berge}, the relative error of the VA being $\approx 7\%$. It is shown
below in Fig. \ref{fig3}(c) that the numerically found counterpart of the
variational value (\ref{cubic-VA}) is%
\begin{equation}
\left( N_{\mathrm{Townes}}\right) _{\mathrm{num}}\approx 1.23.
\label{cubic-num}
\end{equation}%
Thus, the relative error of the VA prediction (\ref{cubic-VA}) is $\approx
13\%$. A relatively large size of the error is a consequence of the
complicated structure of the equation with the fractional diffraction,
especially for values of $\alpha $ which are taken far from the
normal-diffraction limit, $\alpha =2$.

The VA also makes it possible to predict if dependence $N(\alpha )$ for a
fixed value of $\mu $ is growing or decaying. To make this conclusion, one
should take into account that, at $\alpha =2$, the usual NLSE with the cubic
nonlinearity gives rise to the following $N(\mu )$ dependences for the
commonly known soliton solutions:%
\begin{equation}
N_{\alpha =2}=2\sqrt{-2\mu }.  \label{cubic-usual}
\end{equation}%
Comparing it with the variational value (\ref{cubic-VA}), one concludes that
the $N(\alpha )$ dependence is \emph{growing} at $-\mu >-\mu _{\mathrm{crit}%
}\approx 1/4$, and \emph{decaying} at $-\mu <-\mu _{\mathrm{crit}}$. This
prediction agrees with numerical results displayed below in Fig. \ref{fig3}%
(c).

\subsection{Numerical findings}

\begin{figure}[tbp]
\begin{center}
\includegraphics[width=0.9\columnwidth]{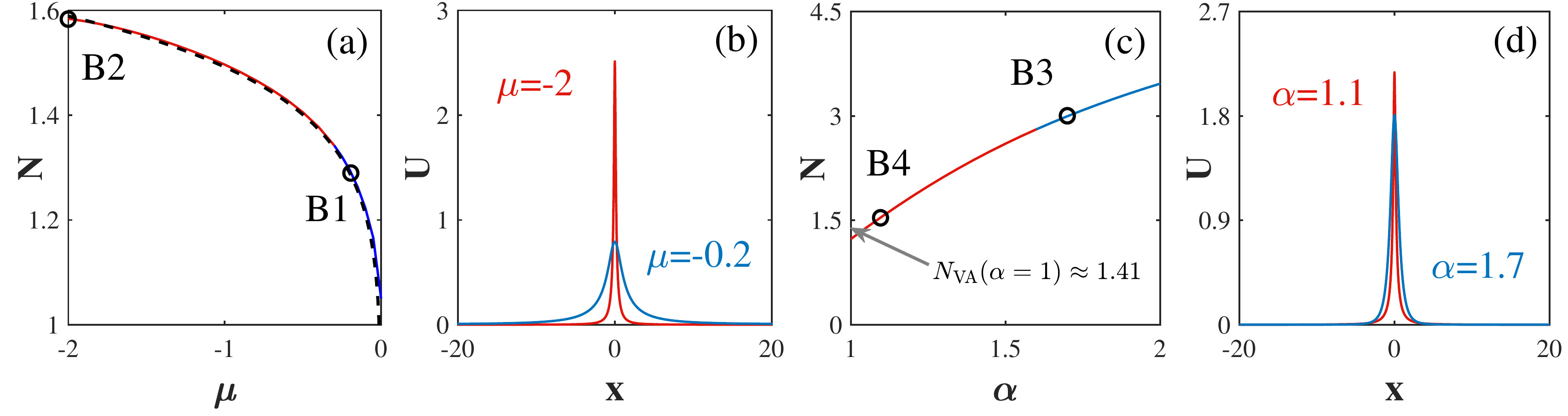}
\end{center}
\caption{(a) Dependence $N(\protect\mu )$ and (b) profiles of solitons
labeled B1 and B2 in (a) (corresponding to $\protect\mu =-0.2$ and $-2$,
respectively), as produced by numerical solution of Eq. (\protect\ref{NLFSES}%
) in the free space ($V=0$), for a fixed value of the L\'{e}vy index, $%
\protect\alpha =1.1$, and nonlinearity coefficient, $g=1$. (c) Dependence $N(%
\protect\alpha )$ and (d) profiles of solitons labeled B3 and B4 in (c)
(corresponding to $\protect\alpha =1.7$ and $1.1$, respectively) for a fixed
propagation constant, $-\protect\mu =1.5$, and $g=1$. Blue and red segments
in panels (a) and (c) denote subfamilies of stable and unstable solitons,
respectively. The black dashed line in (a) represents the analytical scaling
relation (\protect\ref{cubic}). The value marked by the arrow is the $%
\protect\mu $-independent one (\protect\ref{cubic-VA}), predicted by the VA
for the degenerate family of the \textit{quasi-Townes solitons}. Its
numerically found counterpart is given by Eq. (\protect\ref{cubic-num}). The
stability and evolution of the solitons labeled by B1--B4 are displayed in
Fig. \protect\ref{fig4}. The plots are borrowed from Ref. \protect\cite%
{Liangwei} (unpublished).}
\label{fig3}
\end{figure}

Generic numerically obtained results for soliton families produced by Eqs. (%
\ref{NLFSE}) and (\ref{NLFSES}) are displayed in Figs. \ref{fig3} and \ref%
{fig4}. The families are characterized by the dependence of the integral
power $N$, defined as per Eq. (\ref{P}), on the propagation constant $-\mu $
for fixed values of L\'{e}vy index $\alpha $, see a characteristic example
for $\alpha =1.1$ in Fig. \ref{fig3}(a). Note that this value of $\alpha $
is taken far from the normal-diffraction limit, $\alpha =2$, and close to
the collapse boundary, $\alpha =1$, in order to demonstrate the setting in
which the fractional character of the diffraction is essential. It is seen
that the numerically computed dependence exactly follows the analytical
prediction given by Eq. (\ref{cubic}), with a properly fitted constant $%
N_{0}(\alpha )$. Blue and red segments in the $N(\mu )$ curve identify,
respectively, stable and unstable soliton subfamilies.

Further, a curve representing a typical dependence $N(\alpha )$ for a fixed
propagation constant, $-\mu =1.5$, which is also split in stable and
unstable segments, is exhibited in Fig. \ref{fig3}(c). Unlike the $N(\mu )$
dependence, this one cannot be predicted in an exact analytical form.
Nevertheless, as mentioned above, the VA based on ansatz (\ref{ans})
predicts the approximate value given by Eq. (\ref{cubic-VA}) for the
degenerate ($\mu $-independent) norm of the quasi-Townes solitons at $\alpha
=1$, which is close enough to its numerical counterpart (\ref{cubic-num}).
At $\alpha =2$, the numerical value $N_{\mathrm{numer}}\left( \mu
=1.5,\alpha =2\right) \approx 3.46$ is in complete agreement with the exact
analytical value given by Eq. (\ref{cubic-usual}) (it is $N(\mu =-1.5)=2%
\sqrt{3}$).

Typical profiles of the solitons with different values of $\mu $ and $\alpha
$, taken at points labeled B1--B4 in Figs. \ref{fig3}(a) and (c), are
presented in Figs. \ref{fig3}(b,d), respectively. In particular, the trend
of the solitons to get narrower with the increase of $|\mu |$ is another
manifestation of the scaling expressed by Eq. (\ref{cubic}).

\begin{figure}[tbp]
\begin{center}
\includegraphics[width=0.9\columnwidth]{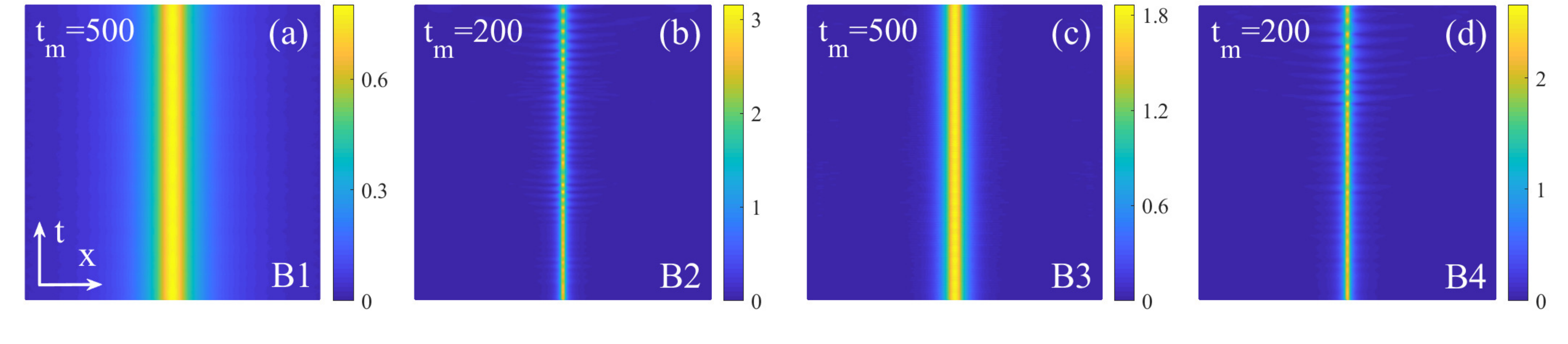}
\end{center}
\caption{The perturbed evolution of the solitons corresponding to labels
B1--B4 in Figs. \protect\ref{fig3}(a,c). The respective values of the
parameters are $\protect\alpha =1.1$, $g=1$, $\protect\mu =-0.2$ (a); $%
\protect\alpha =1.1$, $\protect\mu =-2$ (b); $\protect\alpha =1.7$, $\protect%
\mu =-1.5$ (c); $\protect\alpha =1.1$, $\protect\mu =-1.5$ (d). The
evolution is plotted in the spatial domain $-10<x<+10$. Values $\mathrm{t}_{%
\mathrm{m}}$ indicate intervals of the scaled propagation distance in the
respective panels, $0<z<\mathrm{t}_{\mathrm{m}}$. In panels (a) and (c),
larger $\mathrm{t}_{\mathrm{m}}$ is taken to corroborate the full stability
of the solitons. The plots are borrowed from Ref. \protect\cite{Liangwei}
(unpublished).}
\label{fig4}
\end{figure}

Stable and unstable perturbed propagation of the solitons whose stationary
shapes are shown in Figs. \ref{fig3}(b,d) is displayed in Fig. \ref{fig4}.
It is seen that, if the solitons are unstable, dynamical manifestations of
the instability are quite weak, in the form of spontaneously developing
small-amplitude intrinsic vibrations of the solitons. The instability of the
solitons belonging to the red segments of the $N(\mu )$ and $N(\alpha )$
dependences in Figs. \ref{fig3}(a) and (c) is always accounted for by a pair
of real eigenvalues $\pm \lambda $ produced by numerical solution of Eqs. (%
\ref{LAS}).

\section{Further results for nonlinear modes in one-dimensional fractional
waveguides}

\subsection{Systems with trapping potentials}

While the possibility of the collapse in fractional NLSE (\ref{NLFSE}) gives
rise to instability of all solitons at $\alpha \leq 1$ in the free space
(with $V=0$), the parabolic trapping potential,
\begin{equation}
V(x)=\left( \Omega ^{2}/2\right) x^{2},  \label{V}
\end{equation}%
helps to create stable solitons even in this case \cite{Frac9}. This finding
is similar to the well-known fact that the parabolic trap lifts the norm
degeneracy of the Townes solitons and stabilizes their entire family against
the critical collapse in the framework of the usual two-dimensional NLSE
with the cubic self-attraction \cite{Alexander,PLA}. Moreover, the same
potential makes it possible to predict the existence of stable higher-order
(multi-peak) solutions of Eq. (\ref{NLFSE}). Such solutions may be
considered as a nonlinear extension of various excited bound states
maintained by the parabolic trapping potential in the linear Schr\"{o}dinger
equation. The latter finding is similar to the ability of the 2D parabolic
potential to partly stabilize a family of trapped vortex solitons with
winding number $n=1$ (i.e., the lowest-order excited states in 2D), in the
framework of the cubic self-attractive NLSE \cite{Alexander,PLA}.

\begin{figure}[tbp]
\begin{center}
\includegraphics[width=0.72\columnwidth]{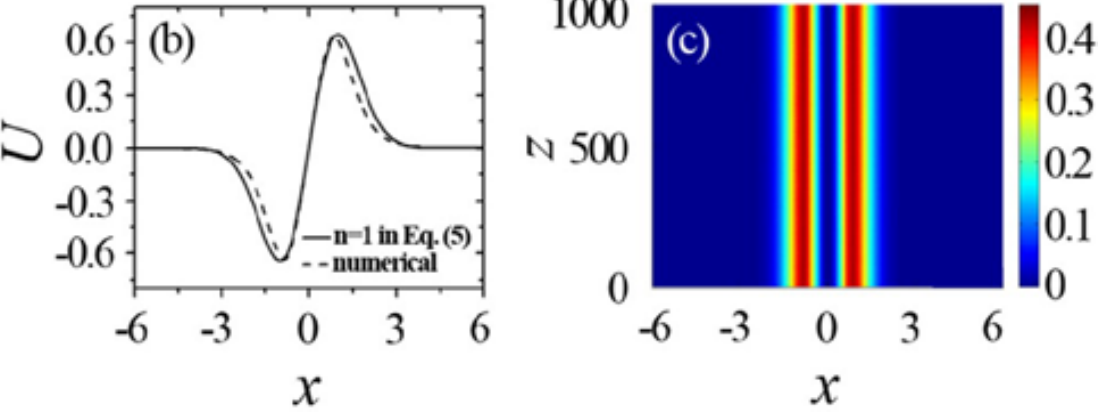}
\end{center}
\caption{An example of a stable dipole-mode soliton produced by Eq. (\protect
\ref{NLFSE}) with L\'{e}vy index $\protect\alpha =1$, parabolic trapping
potential (\protect\ref{V}) with $\Omega ^{2}=1$, i.e., $V(x)=0.5x^{2}$, and
nonlinearity strength $g=0.5$. The propagation constant of this state is $-%
\protect\mu =1.16$. The stationary profile of the soliton and its
propagation, in terms of the local-power distribution, are displayed in the
left and right panels, respectively The solid line in the left panel
represents the analytical fit provided by Eq. (\protect\ref{fit}) with $n=1$%
. Reprinted with permission from Ref. \protect\cite{Frac9}. Copyright 2020
Elsevier.}
\label{fig5}
\end{figure}

The higher-order solitons of orders $n=1,2,3,...$ , supported by Eq. (\ref%
{NLFSES}) with potential (\ref{V}), may be approximated by the commonly
known stationary wave functions of eigenstates of the quantum-mechanical
harmonic oscillator (corresponding to $\alpha =2$),%
\begin{equation}
U(x)=A\exp \left( -\frac{\Omega }{2}x^{2}\right) H_{n}(x),  \label{fit}
\end{equation}%
where $H_{n}(x)$ are Hermite polynomials, and $A$ is an amplitude. An
example of a stable dipole-mode soliton, corresponding to $n=1$, and its fit
provided by Eq. (\ref{fit}) with properly chosen $A$, is displayed in Fig. %
\ref{fig5} for $\alpha =1$, which corresponds to the limit of the
collapse-induced instability in the free space.

\begin{figure}[tbp]
\begin{center}
\includegraphics[width=0.75\columnwidth]{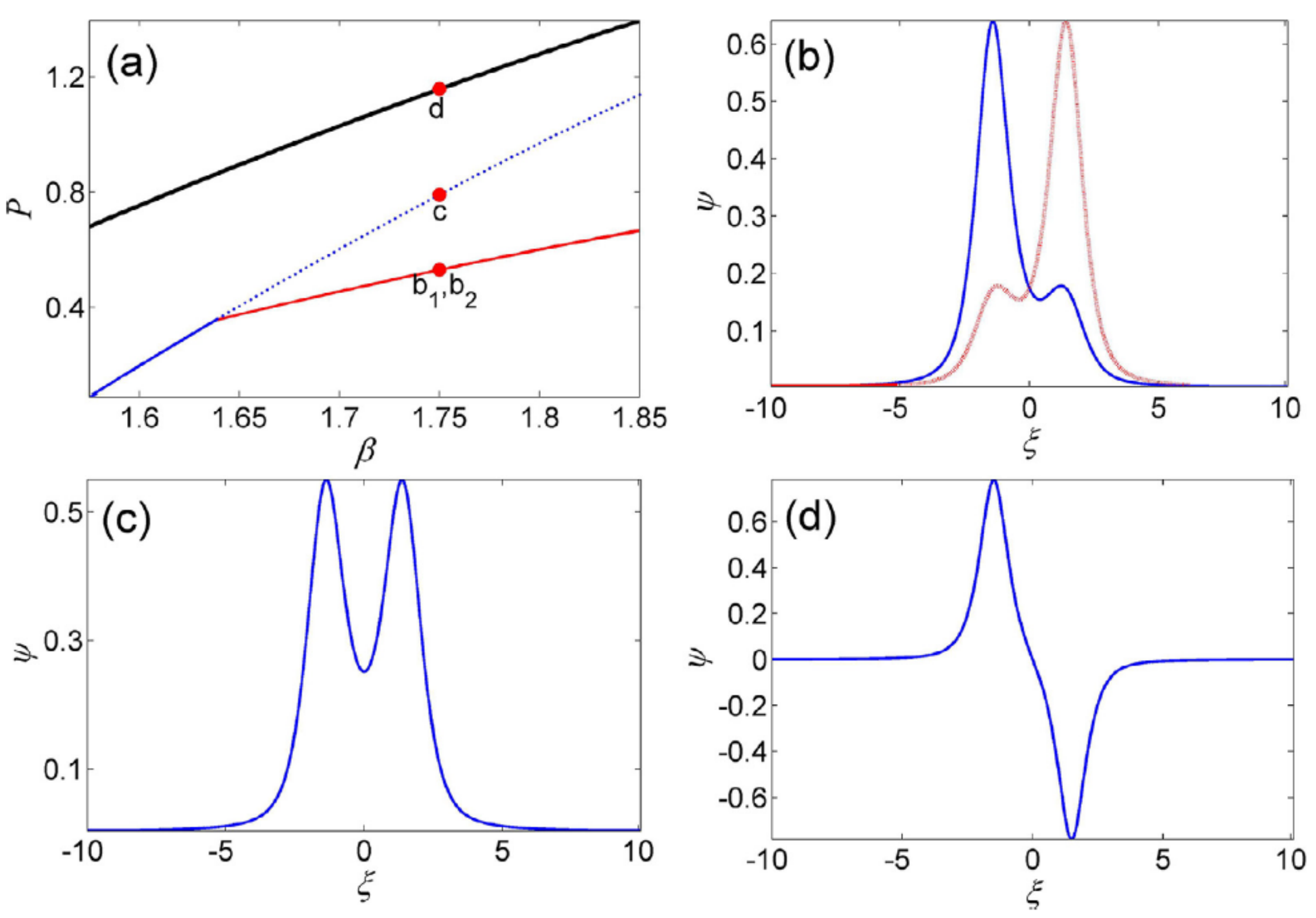}
\end{center}
\caption{(a) Blue, red, and black lines show the norm (alias power, here
denoted $P$) of symmetric, asymmetric, and antisymmetric solitons vs. their
propagations constants (denoted $\protect\beta \equiv -\protect\mu $ here)
in the fractional system based on Eq. (\protect\ref{NLFSE}) with $g=1$ (the
self-focusing nonlinearity) and the double-well potential (\protect\ref%
{double}). The solid and dotted segments of the blue line represent,
respectively, stable and unstable subfamilies of symmetric solitons, below
and above the symmetry-breaking bifurcation. The branches of antisymmetric
and asymmetric solitons are completely stable. Here and in Figs. \protect\ref%
{fig7} and \protect\ref{fig8} parameters of potential (\protect\ref{double})
are $V_{0}=1$, $w=1.4$, $x_{0}=1.5$. Panels (b), (c), and (d) display,
severally, a pair of stable asymmetric solitons (mirror images of each
other), an unstable symmetric soliton, and a stable antisymmetric one, all
taken at $\protect\beta =1.75$ (these solitons correspond to dots b$_{1,2}$,
c, and d in panel (a)). Reprinted with permission from Ref. \protect\cite%
{Frac17}. Copyright 2020 Elsevier.}
\label{fig6}
\end{figure}

\begin{figure}[tbp]
\begin{center}
\includegraphics[width=0.75\columnwidth]{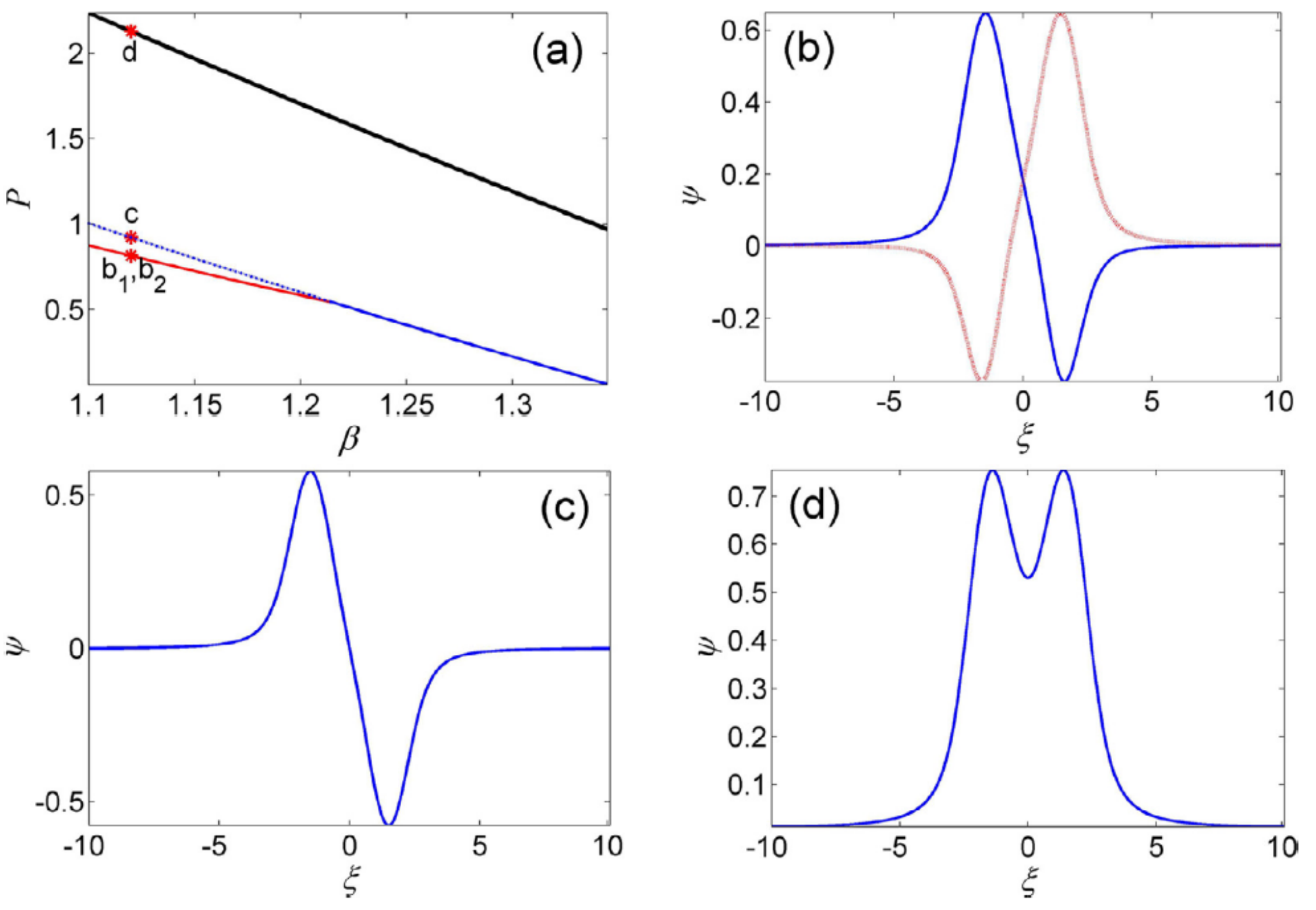}
\end{center}
\caption{(a) Blue, red, and black lines show, respectively, the norm (power,
here denoted $P$) of antisymmetric, asymmetric, and symmetric solitons vs.
their propagations constants (denoted $\protect\beta \equiv -\protect\mu $
here) in the fractional system based on Eq. (\protect\ref{NLFSE}) with the
double-well potential (\protect\ref{double}) and $g=-1$ (the self-defocusing
nonlinearity). Other parameters are the same as in Fig. \protect\ref{fig6}.
The solid and dotted segments of the blue line represent, respectively,
stable and unstable subfamilies of antisymmetric solitons, below and above
the symmetry-breaking bifurcation. The branches of antisymmetric and
asymmetric solitons are completely stable. Panels (b), (c), and (d) display,
severally, a pair of stable asymmetric solitons (mirror images of each
other), an unstable antisymmetric soliton, and a stable symmetric one, all
taken at $\protect\beta =1.12$ (these solitons correspond to dots b$_{1,2}$,
c, and d in panel (a)). Reprinted with permission from Ref. \protect\cite%
{Frac17}. Copyright 2020 Elsevier.}
\label{fig7}
\end{figure}

Another noteworthy effect induced by the potential term in the fractional
NLSE was recently considered in Ref. \cite{Frac17}, which addressed Eq. (\ref%
{NLFSE}) with a \textit{double-well potential}, \textit{viz}.,
\begin{equation}
V(x)=-V_{0}\left[ \exp \left( -\left( \frac{x+x_{0}}{w}\right) ^{2}\right)
+\exp \left( -\left( \frac{x-x_{0}}{w}\right) ^{2}\right) \right] ,
\label{double}
\end{equation}%
with $V_{0}>0$ and width $w$. Equation (\ref{NLFSE}) with this potential may
support solitons which are symmetric or antisymmetric with respect to the
two potential wells. An effect previously studied in detail in the framework
of the usual NLSE (with $\alpha =2$) is that, when the soliton's power
exceeds a critical value, i.e., the nonlinearity is strong enough, symmetric
solitons become unstable and are replaced by stable asymmetric ones, in the
case of the self-focusing nonlinearity, i.e., $g>0$ in Eq. (\ref{NLFSE}),
while antisymmetric solitons remain stable with the increase of their norm
\cite{book}. Alternatively, an antisymmetry-breaking bifurcation takes place
with antisymmetric solitons (while the symmetric ones remain stable) when
the soliton's norm exceeds a critical value in the case of the
self-defocusing nonlinearity, corresponding to $g<0$ in Eq. (\ref{NLFSE}).
Such symmetry-- and antisymmetry-breaking bifurcations\textit{\ }(phase
transitions) in the fractional NLSE with potential (\ref{double}) are shown,
severally, in Figs. \ref{fig6} and \ref{fig7} for the L\'{e}vy index $\alpha
=1.1$, which makes the system essentially different from the usual NLSE,
corresponding to $\alpha =2$.

\begin{figure}[tbp]
\begin{center}
\subfigure[]{\includegraphics[width=0.24\textwidth]{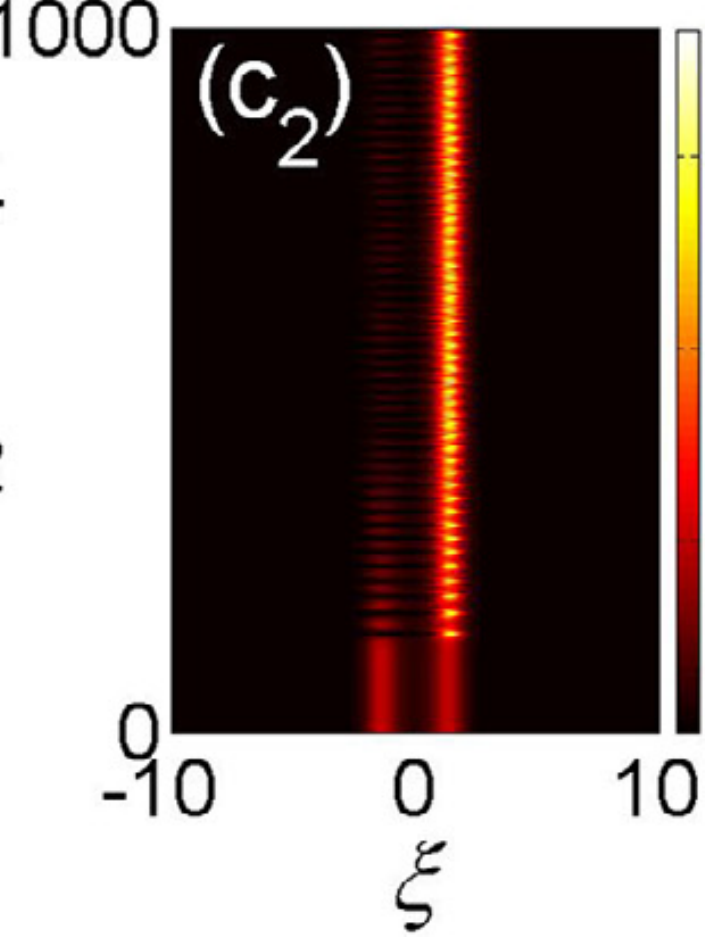}}%
\subfigure[]{\includegraphics[width=0.24\textwidth]{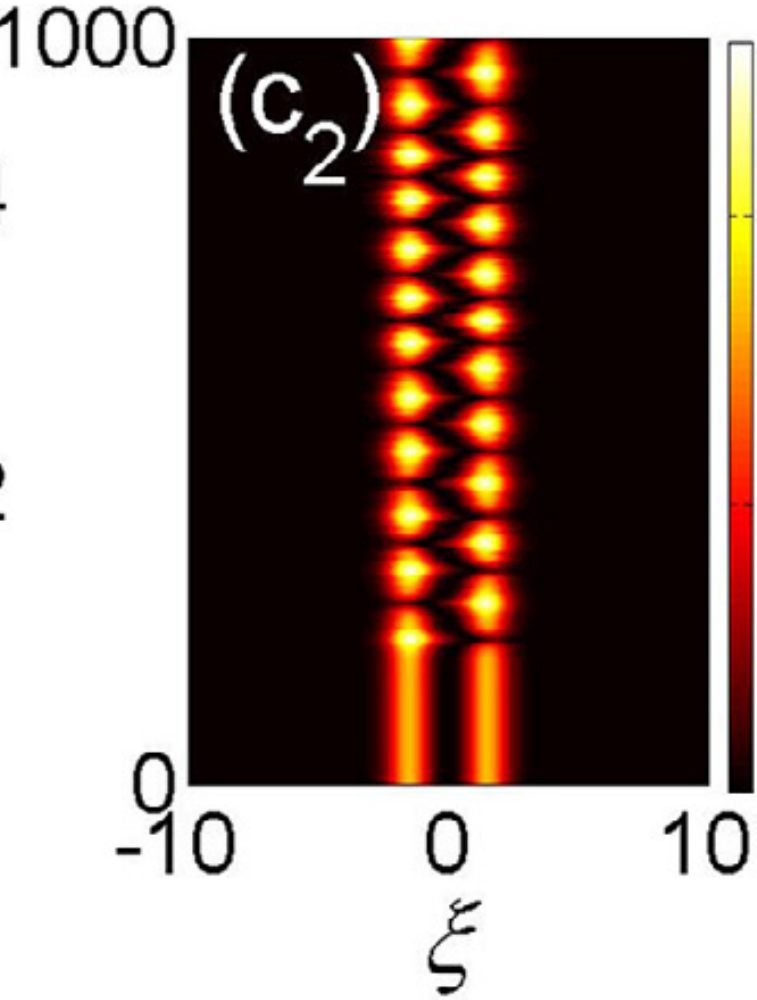}}
\end{center}
\caption{ The perturbed evolution of unstable symmetric (a) and
antisymmetric (b) solitons, produced by simulations of Eq. (\protect\ref%
{NLFSE}) with the double-well potential (\protect\ref{double}), L\'{e}vy
index $\protect\alpha =1.1$, and nonlinearity strength $g=+1$ (a) and $-1$
(b). The propagation constants of the unstable symmetric and antisymmetric
solitons are $\protect\mu =-1.62$ and $\protect\mu =-1.12$, respectively.
Reprinted with permission from Ref. \protect\cite{Frac17}. Copyright 2020
Elsevier.}
\label{fig8}
\end{figure}

As shown in Fig. \ref{fig8}(a), simulations of the perturbed evolution,
performed for the present model in Ref. \cite{Frac17}, demonstrate that an
unstable symmetric soliton (in the case of the self-attractive nonlinearity)
breaks its symmetry and spontaneously transforms into a dynamical state
which is close to either one of stable asymmetric solitons existing with the
same power (a chiral pair of such solitons is shown in Fig. \ref{fig6}(b)).
On the other hand, simulations of the evolution of unstable antisymmetric
solitons (in the case of the self-repulsion) produce a different result, as
shown in Fig. \ref{fig8}(b): the soliton does not tend to spontaneously
transform into either one of the mutually chiral asymmetric solitons, but
instead oscillates between them.

Furthermore, the action of the trapping potential (\ref{double}), even if it
is essentially weaker than the parabolic potential (\ref{V}), provides an
effect similar to the above-mentioned one induced by potential (\ref{V}),
\textit{viz}., stabilization of symmetric, antisymmetric, and asymmetric
solitons in the case of $\alpha \leq 1$, when all solitons are unstable in
the free space, due of the occurrence of the critical ($\alpha =1$) or
supercritical ($\alpha <1$) collapse. In particular, it was demonstrated
that the symmetry-breaking bifurcation, quite similar to the one displayed
in Fig. \ref{fig6} (thus, including \emph{stable} soliton branches) takes
place in the same system at $\alpha =0.8$ \cite{Frac17}.

\subsection{Dissipative solitons produced by the fractional complex
Ginzburg-Landau equation (CGLE)}

An essential extension of the fractional NLSE was proposed in Ref. \cite%
{Frac14}, in the form of the equation with complex coefficients (including
the coefficient in front of the fractional-diffraction term), i.e., the
fractional CGLE. It models the propagation of light in waveguides which, in
addition to the fractional diffraction, include losses and gain:
\begin{equation}
i\frac{\partial \Psi }{\partial z}=-i\delta \Psi +\left( \frac{1}{2}-i\beta
\right) \left( -\frac{\partial ^{2}}{\partial x^{2}}\right) ^{\alpha /2}\Psi
+\left( i\varepsilon -1\right) \left\vert \Psi \right\vert ^{2}\Psi +\left(
-i\mu +\nu \right) |\Psi |^{4}\Psi ,  \label{quintic}
\end{equation}%
where $\delta >0$, $\mu >0$, and $\beta \geq 0$ account for, respectively,
the linear, quintic, and diffraction losses (fractional diffusion), $%
\varepsilon >0$ is the cubic gain, while $\nu \geq 0$ represents quintic
self-defocusing, if it is present in the medium, and the cubic self-focusing
term is the same as in Eq. (\ref{NLFSE}) with $g=1$.

\begin{figure}[tbp]
\begin{center}
\subfigure[]{\includegraphics[width=0.40\textwidth]{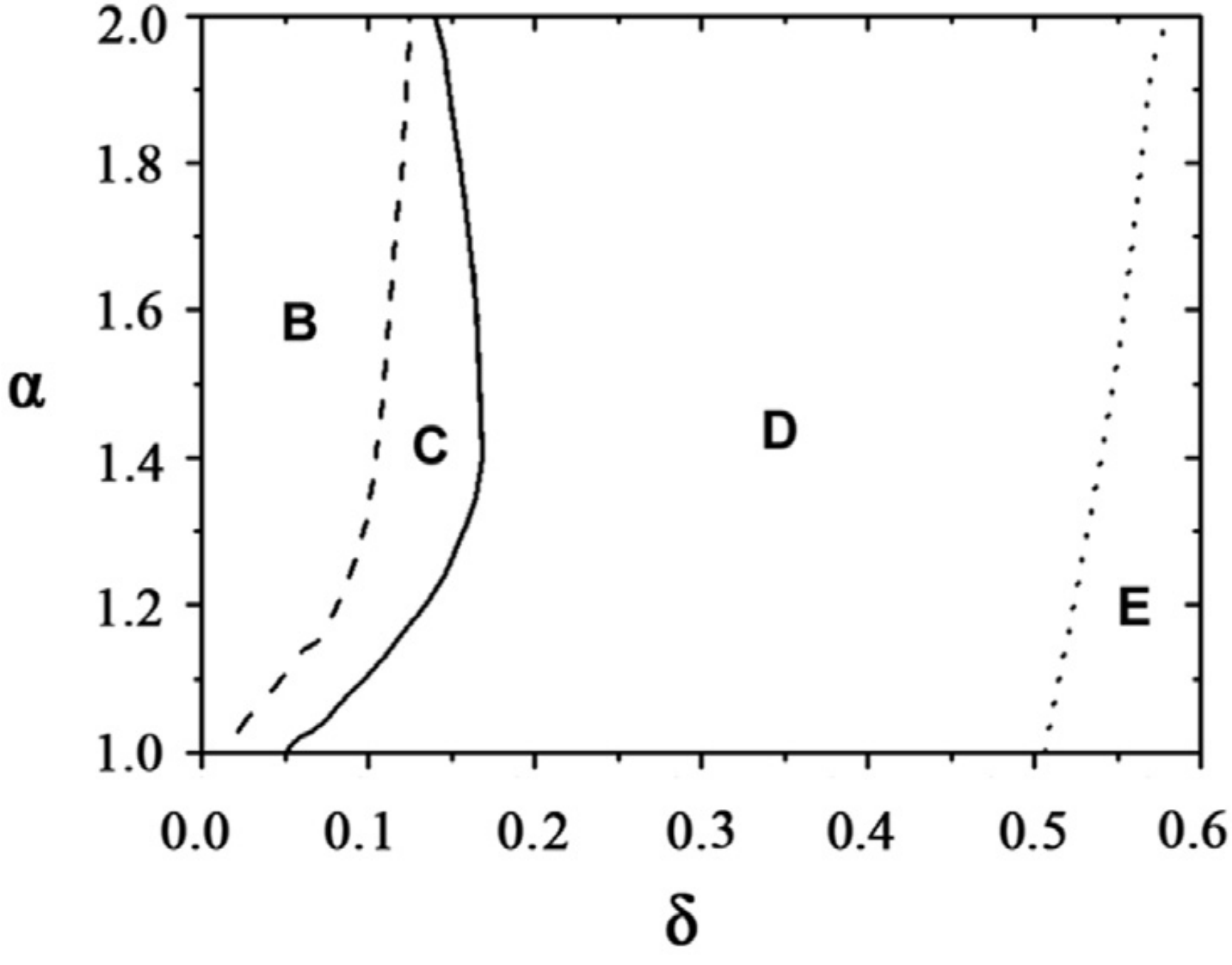}}%
\subfigure[]{\includegraphics[width=0.37\textwidth]{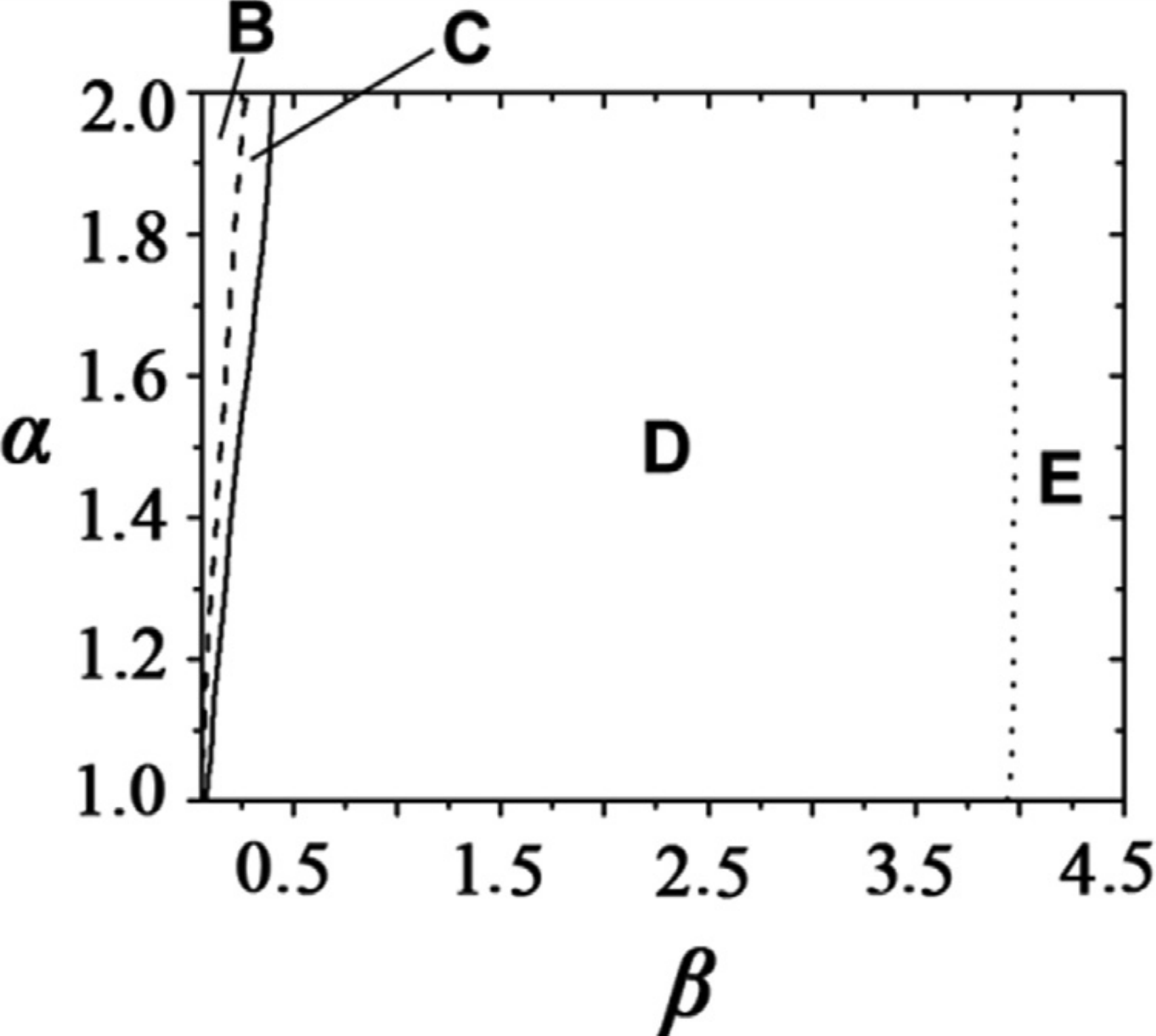}}
\end{center}
\caption{Charts of different established states produced by simulations of
the fractional CGLE with the cubic-quintic nonlinearity (\protect\ref%
{quintic}), plotted in planes of the most essential parameters: linear loss $%
\protect\delta $ and L\'{e}vy index $\protect\alpha $ ((a), with fixed $%
\protect\beta =0.1$), or the fractional-diffusion coefficient $\protect\beta
$ and $\protect\alpha $ ((b), with $\protect\delta =0.1$). Other
coefficients are $\protect\varepsilon =1.7$, $\protect\mu =1$, and $\protect%
\nu =0.115$. In the underdamped (low-loss) parameter area B, a uniform state
extends, in direct simulations, to occupy the entire spatial domain. In
areas C and D, stable dissipative solitons emerge (directly in D, and via an
initial unstable-evolution stage in C). The input decays to zero in the
overdamped area E. Reprinted with permission from Ref. \protect\cite{Frac14}%
. Copyright 2020 Elsevier.}
\label{fig9}
\end{figure}

Equation (\ref{quintic}) gives rise to dissipative solitons. Unlike those in
the conservative NLSE, dissipative solitons exist not in families
parametrized by the propagation constant, but as isolated localized
solutions, whose stability is the major issue \cite{Frac14}. Systematic
simulations of Eq. (\ref{quintic}) have produced charts of different
dynamical states, which are displayed in Fig. \ref{fig9}. The charts
demonstrate a broad parameter region in which stable dissipative solitons
appear.

\begin{figure}[tbp]
\begin{center}
\includegraphics[width=0.7\columnwidth]{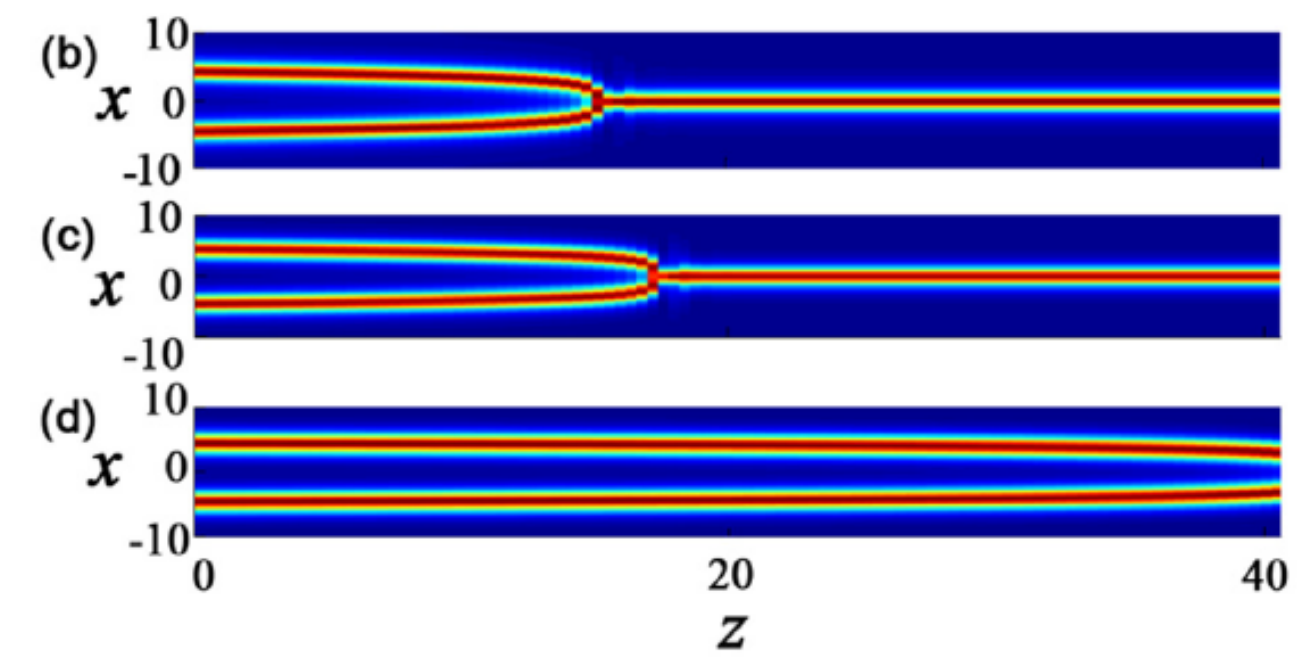}
\end{center}
\caption{The merger of initially separated stable dissipative solitons, with
zero phase difference between them, as produced by simulations of the
fractional cubic-quintic CGLE (\protect\ref{quintic}), with parameters $%
\protect\delta =0.3$, $\protect\beta =0.1$, $\protect\varepsilon =1.7$, $%
\protect\nu =0.115$, $\protect\mu =1$, and different values of the L\'{e}vy
index: $\protect\alpha =1.5$ (b), $1.8$ (c), and $2.0$ (d). Reprinted with
permission from Ref. \protect\cite{Frac14}. Copyright 2020 Elsevier.}
\label{fig10}
\end{figure}

It is worthy to note that interaction between stable in-phase dissipative
solitons, initially separated by some distance, leads to their merger into a
single one, as shown in Fig. \ref{fig10}. It is seen that the merger is
extremely slow in the case of the usual diffraction ($\alpha =2$), while the
fractional diffraction essentially accelerates the process, due to the fact
that the respective operator (\ref{FracDefi}) actually makes the interaction
between the separated solitons nonlocal, i.e., stronger.

\section{Vortex modes in two-dimensional (2D) fractional-diffraction settings%
}

\subsection{Stationary vortex solitons}

As concerns the propagation of light beams with a 2D transverse structure in
fractional media, the transmission of Airy waves, including ones carrying
intrinsic vorticity (the optical angular momentum), was analyzed in the
framework of the linear variant of Eq. (\ref{2+1D}) with $g=V=0$ \cite%
{Yingji1,Yingji2}. In particular, it is worthy to note that the tightest
self-focusing of the Airy-ring input is attained at the value of the L\'{e}%
vy index $\alpha \approx 1.4$, which is essentially different from $\alpha
=2 $ corresponding to the normal 2D diffraction.

Dynamics of vortex solitons and vorticity-carrying ring-shaped soliton
clusters was recently addressed in Refs. \cite{Frac6,Frac10}, and \cite%
{soliton5} in the framework of the fractional NLSE with the cubic-quintic
nonlinearity:
\begin{equation}
i\frac{\partial \Psi }{\partial z}=\frac{1}{2}\left( -\frac{\partial ^{2}}{%
\partial x^{2}}-\frac{\partial ^{2}}{\partial y^{2}}\right) ^{\alpha /2}\Psi
-\left\vert \Psi \right\vert ^{2}\Psi +|\Psi |^{4}\Psi ,  \label{CQ}
\end{equation}%
cf. Eq. (\ref{quintic}). The 2D fractional-diffraction operator in Eq. (\ref%
{CQ}) is defined as per Eq. (\ref{2D operator}).

Stationary solutions to Eq. (\ref{CQ}), with propagation constant $-\mu >0$,
integer vorticity (alias winding number) $s$, and real amplitude function $%
\Phi (r)$, are looked for, in polar coordinates $\left( r,\theta \right) $,
as%
\begin{equation}
\Psi =\exp \left( -i\mu z+is\theta \right) U(r).  \label{Phi}
\end{equation}%
Unlike the NLSE with the usual diffraction ($\alpha =2$), the substitution
of ansatz (\ref{Phi}) in Eq. (\ref{CQ}) with the fractional diffraction
operator is not trivial. Nevertheless, using the definition (\ref{2D
operator}) of the operator, the respective calculation can be performed,
demonstrating that the vortex ansatz is compatible with Eq. (\ref{CQ}) (in
other words, the angular-momentum operator commutes with the system's
Hamiltonian), the resulting equation being%
\begin{equation}
\mu U=\frac{1}{2}\left( -\nabla _{r}^{2}\right) ^{\alpha /2}U-U^{3}+U^{5},
\label{U}
\end{equation}%
where the fractional radial Laplacian is obtained in a cumbersome form:%
\begin{equation}
\left( -\nabla _{r}^{2}\right) ^{\alpha /2}U(r)=\frac{1}{2\pi }%
\int_{0}^{\infty }q^{\alpha +1}dq\int_{0}^{\infty }r^{\prime }dr^{\prime
}\int_{0}^{2\pi }d\chi \cos \left( s\chi \right) J_{0}\left( q\sqrt{%
r^{2}+\left( r^{\prime }\right) ^{2}-2rr^{\prime }\cos \chi }\right)
U(r^{\prime }),  \label{radial}
\end{equation}%
with Bessel function $J_{0}$.

Localized solutions of Eq. (\ref{U}) are characterized by the 2D norm,
\begin{equation}
N_{\mathrm{2D}}\equiv \int \int \left\vert \Psi \left( x,y\right)
\right\vert ^{2}dxdy\equiv 2\pi \int_{0}^{\infty }U^{2}(r)rdr.  \label{2D}
\end{equation}%
In the absence of the self-defocusing quintic term, a straightforward
corollary of Eq. (\ref{U}) with the cubic-only nonlinearity is a scaling
relation between the norm and propagation constant, cf. Eq. (\ref{cubic}):%
\begin{equation}
N_{\mathrm{2D}}(\mu )=N_{0}(\alpha )\left( -\mu \right) ^{1-2/\alpha }.
\label{N(mu)}
\end{equation}%
Obviously, at all values $\alpha \leq 2$ (i.e., at all relevant values of
the L\'{e}vy index), this relation \emph{does not} satisfy the VK criterion (%
\ref{VaKo}), that is why the inclusion of the quintic self-defocusing term
is necessary for the stability of the resulting solitons, both fundamental ($%
s=0$) and vortical ones. Furthermore, due to the stabilizing effect of the
quintic term, Eq. (\ref{CQ}) supports stable localized states even in the
case of $\alpha \leq 1$, when the cubic self-focusing alone gives rise to
the collapse in the 1D setting. This stabilizing effect, which is maintained
by the quintic nonlinearity in the free space, may be compared to the
above-mentioned stabilization mechanism, provided by the trapping potential.

\begin{figure}[tbp]
\begin{center}
\includegraphics[width=0.93\columnwidth]{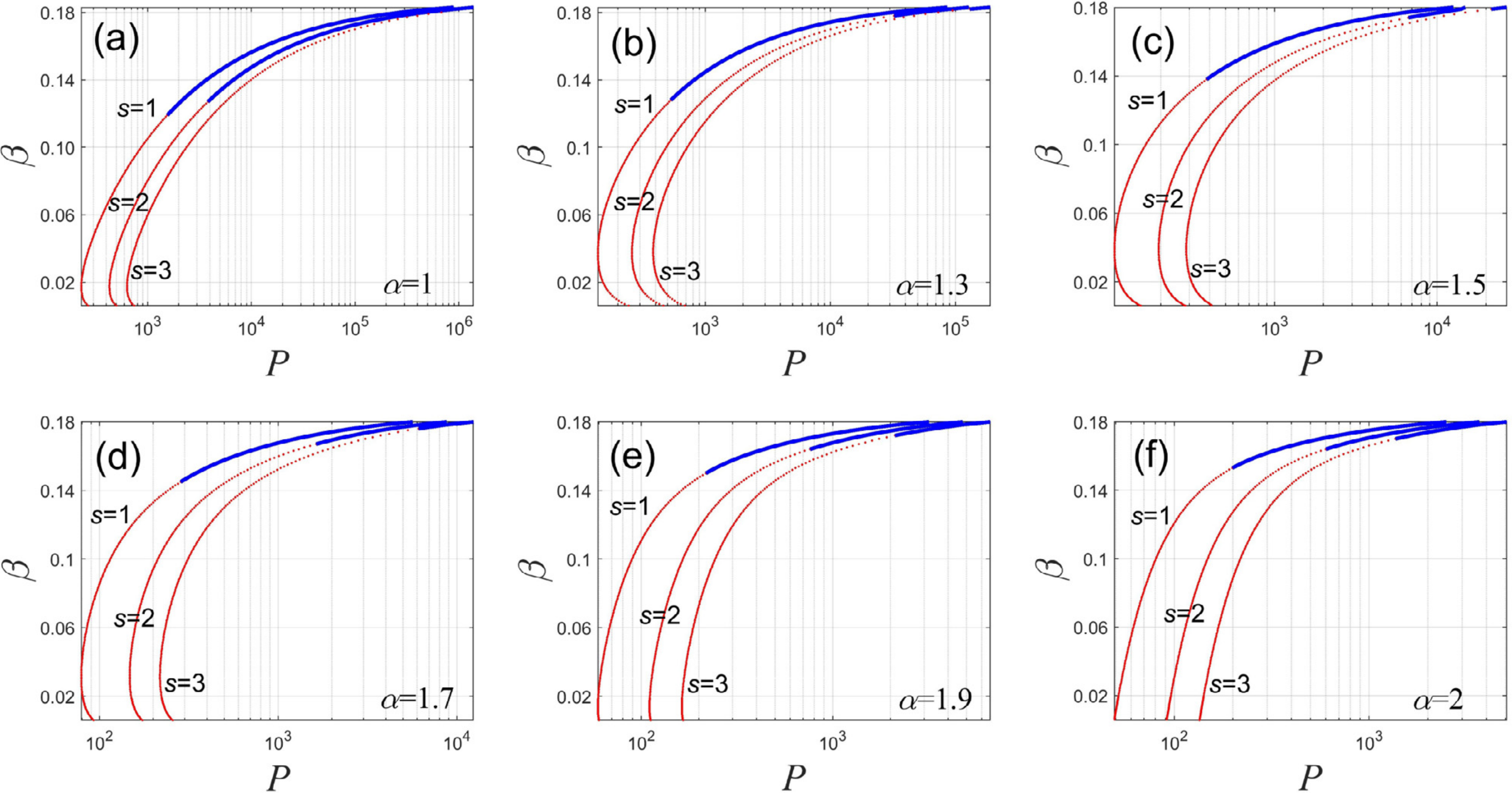}
\end{center}
\caption{The propagation constant, $\protect\beta \equiv -\protect\mu $, vs.
the norm (power) of the 2D solitons with vorticities $s=1,2$, and $3$. The
power is defined here as $P\equiv 2^{2/\protect\alpha }N_{\mathrm{2D}}$, see
Eq. (\protect\ref{2D}). The results are produced by the numerical solution
of Eq. (\protect\ref{U}) for fixed values of the L\'{e}vy index: $\protect%
\alpha =1$ (a), $1.3$ (b), $1.5$ (c), $1.7$ (d), $1.9$ (e), and $2$ (f). The
latter case, corresponding to the usual two-dimensional NLSE, is included
for comparison with the results produced by the fractional diffraction. Note
that values of the power are shown on the logarithmic scale, which is
different in different panels. Blue and red segments represent stable and
unstable vortex states, respectively. Reprinted with permission from Ref.
\protect\cite{Frac6}. Copyright 2020 Elsevier.}
\label{fig11}
\end{figure}

The distribution of the local power in soliton solutions with embedded
vorticity (winding number) $s$ features a ring-like shape (see examples
below in the left column of Fig. \ref{fig13}). Families of such solutions
with $s=1,2,$ and $3$, produced by numerical solution of Eq. (\ref{U}), are
represented by dependences of the 2D norm on the propagation constant, which
are displayed in Fig. \ref{fig11}. These dependences also highlight
relatively small stable subfamilies of the vortex solitons. The stability
was identified by the computation of eigenvalues, using linearized equations
for small perturbations, and verified by direct simulations of the perturbed
evolution \cite{Frac6}.

In the limit of the diverging norm, the propagation constant in all panels
of Fig. \ref{fig11} attains the commonly known limit value for the NLSE with
the cubic-quintic nonlinearity, $\beta _{\max }=3/16$ \cite{Bulgaria}, which
does not depend on the spatial dimension. Further, it is seen in the figure
that the soliton families exist at values of the norm (power) exceeding a
certain threshold value,
\begin{equation}
N\geq N_{\mathrm{thr}}^{(s)}(\alpha ),  \label{thr}
\end{equation}%
although the stability boundary corresponds to much larger values of the
norm. Actually, this feature (i.e., the non-existence of solitons with the
norm falling below a finite threshold value) is well known in models where
the self-focusing nonlinear term, if acting alone, gives rise to the
supercritical collapse, such as the three-dimensional NLSE with the normal
diffraction \cite{old}. Indeed, in collapse-free models solitons with a
vanishingly small norm have a vanishing amplitude and diverging size, the
latter fact implying $\mu \rightarrow -0$. However, Eq. (\ref{N(mu)})
demonstrates that, at any $\alpha <2$, the norm of the solitons is
diverging, rather than vanishing, at $\mu \rightarrow -0$, hence the limit
of $N\rightarrow 0$ cannot be attained.

\begin{figure}[tbp]
\begin{center}
\includegraphics[width=0.47\columnwidth]{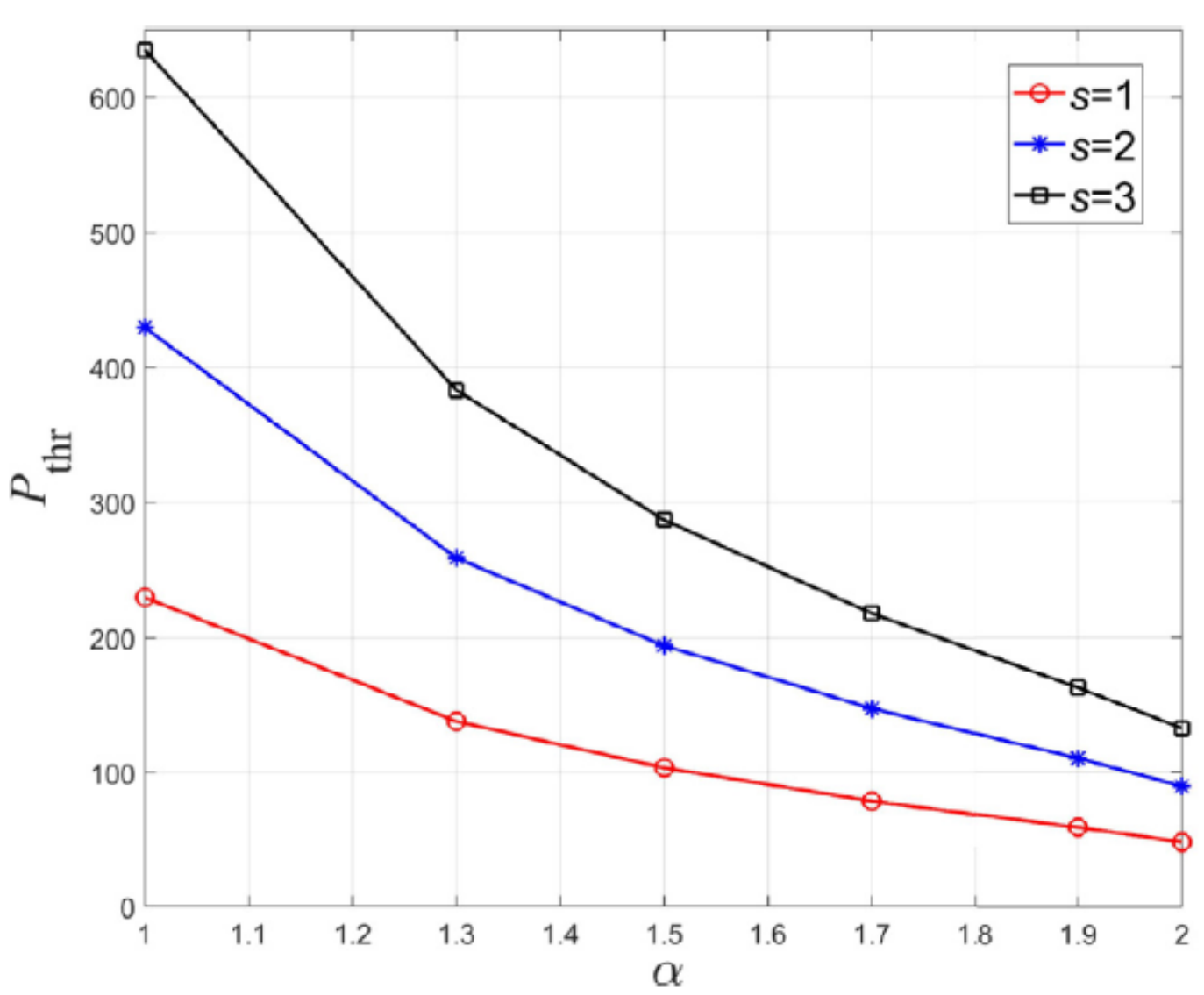}
\end{center}
\caption{The threshold value of the power (norm), $P_{\mathrm{thr}}\equiv
2^{2/\protect\alpha }N_{\mathrm{thr}}^{(s)}(\protect\alpha )$, below which
Eq. (\protect\ref{U}) cannot produce vortex solitons (see Eq. (\protect\ref%
{thr})), vs. the L\'{e}vy index $\protect\alpha $, for vorticities $s=1,2,$
and $3$. In the limit of $\protect\alpha =2$, which corresponds to the
normal (non-fractional) diffraction, the threshold values $N_{\mathrm{thr}%
}^{(s)}(\protect\alpha =2)$ coincide with the norms of the Townes solitons
with the embedded vorticity, see Eq. (\protect\ref{NNN}). Reprinted with
permission from Ref. \protect\cite{Frac6}. Copyright 2020 Elsevier.}
\label{fig12}
\end{figure}

\begin{figure}[tbp]
\begin{center}
\includegraphics[width=0.48\columnwidth]{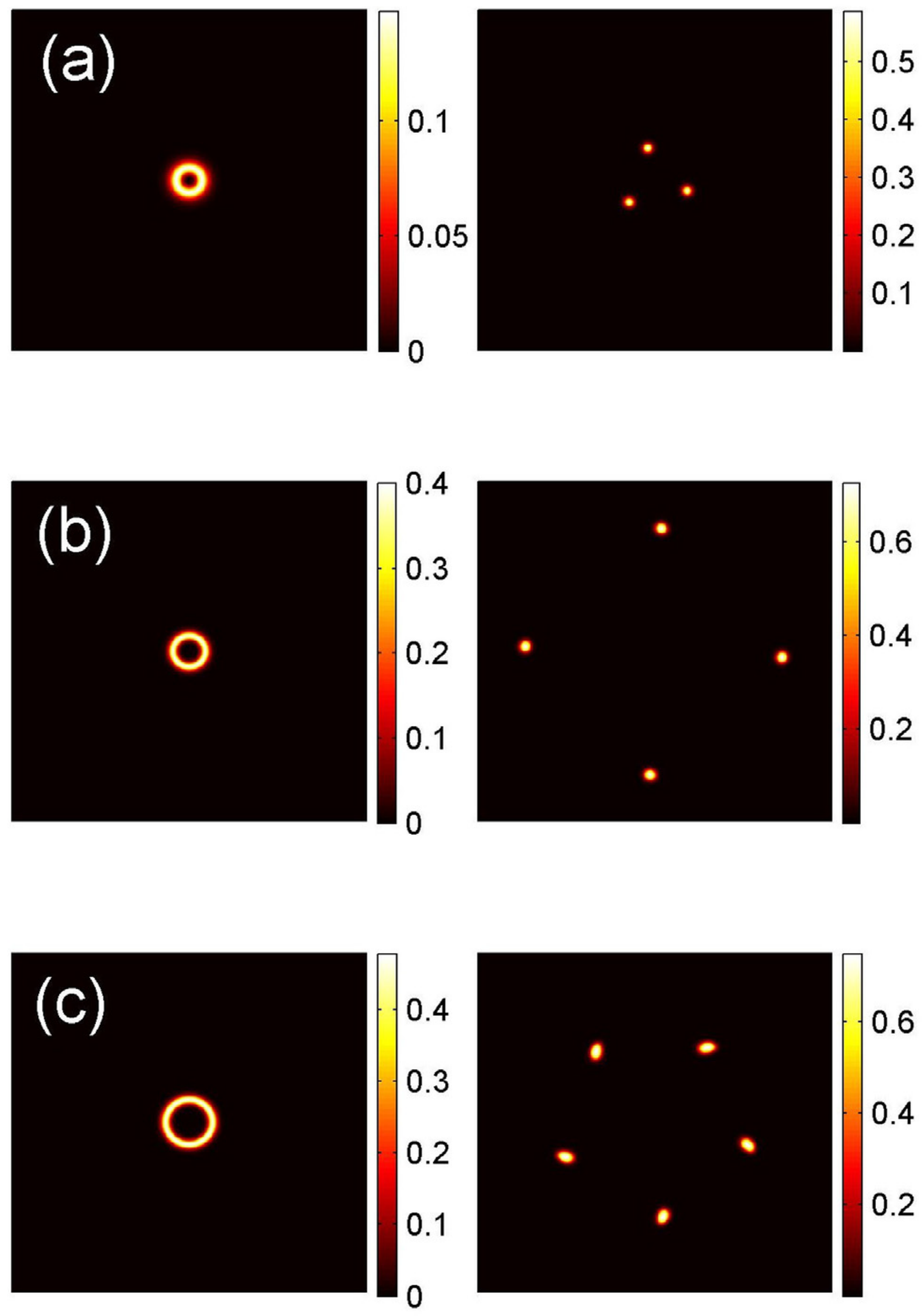}
\end{center}
\caption{Spontaneous splitting of unstable ring-shaped vortex solitons, with
winding numbers $s$ and propagation constants $\protect\mu $, into sets of
fragments, as produced by simulations of Eq. (\protect\ref{2D}) with $%
\protect\alpha =1.5$: (a) $\left( s,\protect\mu =1,-0.03\right) $; (b) $%
\left( 2,-0.08\right) $; (c) $\left( 3,-0.095\right) $. The left and right
panels display, respectively, the local power distribution in the input ($%
z=0 $) and output, taken at $z=500$. All panels show the domain of the $%
(x,y) $ plane of the size $\left( -150,+1.150\right) \times \left(
-150,+1.150\right) $. Reprinted with permission from Ref. \protect\cite%
{Frac6}. Copyright 2020 Elsevier.}
\label{fig13}
\end{figure}

Dependences of the threshold norm on the L\'{e}vy index for $s=1,2,$ and $3$%
\ are displayed in Fig. \ref{fig12}. In the limit of $\alpha =2$, i.e., in
the case of the cubic-quintic NLSE with the normal 2D diffraction, the
respective threshold values $N_{\mathrm{thr}}^{(s)}(\alpha =2)$ do not
vanish either. They correspond to the Townes solitons of the 2D cubic NLSE
with the embedded vorticity \cite{Kruglov,Kruglov2,REV10}, cf. the norm
given by Eq. (\ref{Townes}) for the fundamental Townes solitons (with $s=0$):%
\begin{equation}
N_{\mathrm{thr}}^{(s=1)}(\alpha =2)\approx 24.2;N_{\mathrm{thr}%
}^{(s=2)}(\alpha =2)\approx 44.9;N_{\mathrm{thr}}^{(s=3)}(\alpha =2)\approx
61.3.  \label{NNN}
\end{equation}%
As well as their fundamental counterparts, families of vortex Townes
solitons are degenerate, in the sense that their norm takes the single
value, which depends on $s$ but does not depend on the propagation constant.
Actually, these values may be approximated by an analytical formula obtained
in Ref. \cite{Qin}: $N_{\mathrm{thr}}^{(s)}(\alpha =2)\approx 4\sqrt{3}\pi s$%
.

In simulations of Eq. (\ref{CQ}) those vortex solitons which are unstable
split in sets of separating fragments, as shown in Fig. \ref{fig13}. This is
the usual instability-development scenario for vortex solitons in NLSEs with
diverse nonlinearities \cite{REV10}.

\subsection{Dynamics of vortical clusters}

In addition to the axisymmetric stationary vortex solitons, it is relevant
to consider the dynamics of necklace-shaped clusters. At $z=0$ they are composed as
sets (circular chains) of $M$ fundamental solitons (with zero intrinsic
winding numbers), with equal distances between them,
\begin{equation}
l=2R\sin \left( \pi /M\right) ,  \label{l}
\end{equation}%
and global vorticity, $S=1,2,3,...$, imprinted onto the cluster:%
\begin{gather}
\Psi _{0}\left( x,y\right) =\exp \left( iS\theta \right)
\sum_{m=1}^{M}U_{0}\left( \left\vert \mathbf{r}-\mathbf{r}_{m}\right\vert
\right) ,  \label{input} \\
\mathbf{r}_{m}=R\left\{ \cos \left( \frac{2\pi m}{M}\right) ,\sin \left(
\frac{2\pi m}{M}\right) \right\} .  \label{R}
\end{gather}%
Here $U_{0}\left( \left\vert \mathbf{r}-\mathbf{r}_{m}\right\vert \right) $
is the stationary shape of the fundamental soliton with the center placed at
point $\mathbf{r}=\mathbf{r}_{m}$, and $R$ is the radius of the cluster.

Previously, patterns of the necklace type have drawn much interest in
studies of models based on NLSEs with normal diffraction \cite%
{necklace1,necklace2,necklace3}. In particular, robust soliton necklaces
supported by the cubic-quintic nonlinearity (the same as in Eq. (\ref{CQ}))
were addressed in Ref. \cite{necklace-CQ}.

\begin{figure}[tbp]
\begin{center}
\includegraphics[width=0.82\columnwidth]{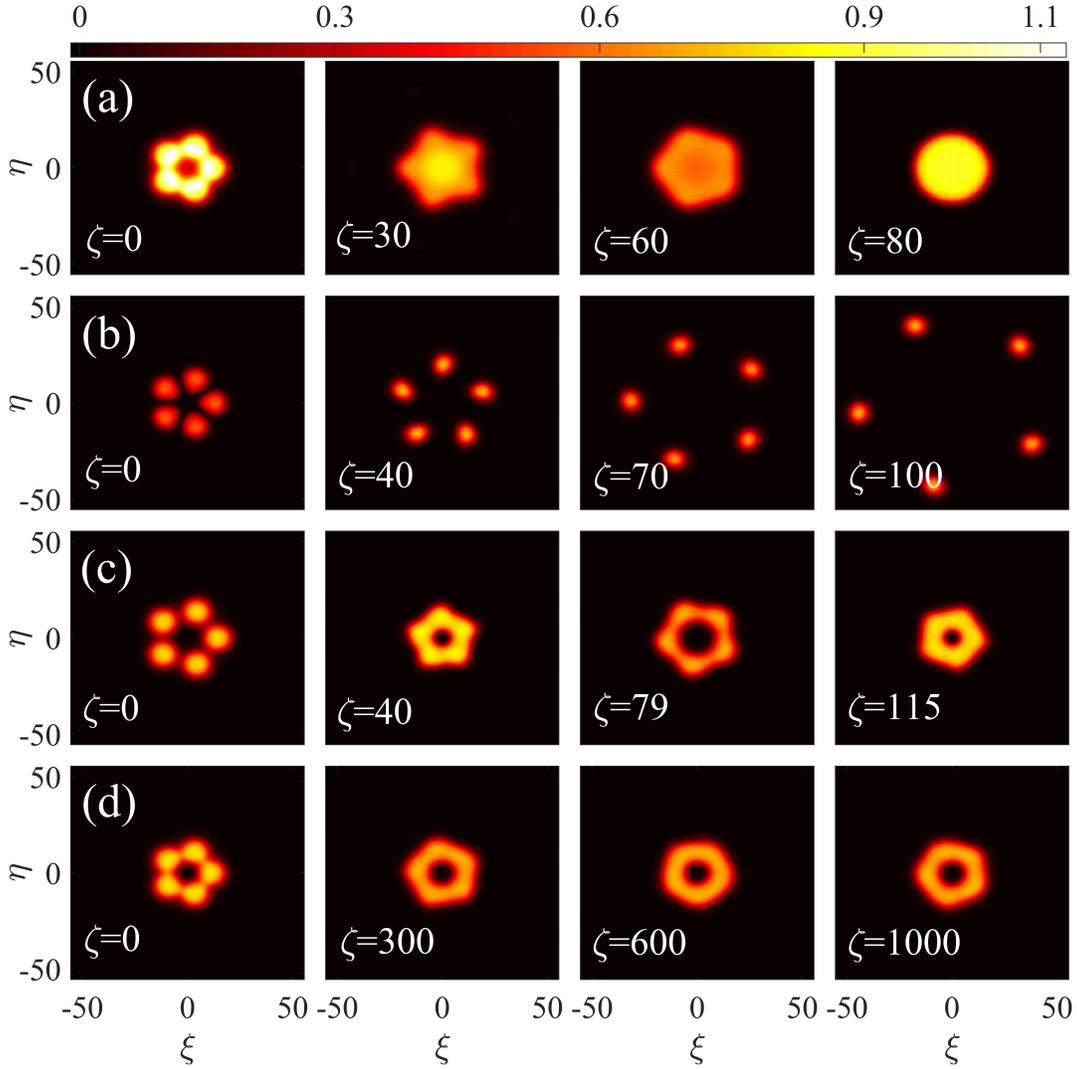}
\end{center}
\caption{The evolution of necklace-shaped clusters initially built, as per
Eqs. (\protect\ref{input}) and (\protect\ref{l}), of $M=5$ fundamental
solitons, carrying overall vorticity $S=0$ (a), $S=2$ (b), and $S=1$ (c,d).
The initial radii of the inputs, and the angular velocity of their rotation,
produced by the evolution, are, respectively, $\left( R,\protect\omega %
)=(11.25,0\right) $ (a); $\left( 10.35,0.0128\right) $ (b); $\left(
13,0.005\right) $ (c); and $\left( 10.35,0.0071\right) $ (d). The results
were produced by simulations of Eq. (\protect\ref{CQ}) with L\'{e}vy index $%
\protect\alpha =1$. In this figure, coordinates $x$, $y$ and $z$ are
denoted, respectively, as $\protect\xi $, $\protect\eta $ and $\protect\zeta
$. The figure is borrowed from Ref. \protect\cite{Frac10}.}
\label{fig14}
\end{figure}

The formation and stability of the circular soliton chains in the framework
of Eq. (\ref{CQ}) was studied in detail in Ref. \cite{Frac10}. Parameters of
the input, taken in the form of Eqs. (\ref{input}) and (\ref{R}), were
chosen so that the axisymmetric vortex soliton with the same vorticity $S$,
and the norm equal to the total norm of the initial soliton chain, is
stable, in terms of Fig. \ref{fig11}. A set of typical results produced by
this input is displayed in Fig. \ref{fig14}. It is seen in row (a) of the
figure that, in the absence of the overall vorticity ($S=0$), the input set
of solitons promptly fuses into a stable fundamental soliton. On the other
hand, if the vorticity is too large, $S\geq 2$, no bound state is formed,
and the initial soliton chain quickly expands, as observed in Fig. \ref%
{fig14}(b).

The formation of quasi-stable slowly rotating necklaces is possible in the
case of $S=1$. In Fig. \ref{fig14}(c), the necklace state features slow
rotation in combination with periodic oscillations in the radial
direction. On the other hand, Fig. \ref{fig14}(d) demonstrates an example of
a rotating necklace with a nearly permanent shape, which remains stable in
the course of very long propagation. Indeed, $z=1000$ in this case is
tantamount, roughly, to $\sim 50$ diffraction (Rayleigh) lengths of the
input pattern, which, in turn, may be estimated as $z_{\mathrm{Rayleigh}%
}\sim \left( 2R\right) ^{\alpha }$.

\subsection{Stabilization of 2D solitons by the trapping potential}

As said above, all soliton solutions of Eq. (\ref{2+1D}) with the cubic
self-focusing nonlinearity ($g=1$) in the free space ($V=0$) are completely
unstable at all values of the L\'{e}vy index, $\alpha \leq 2$. Instead of
the quintic self-defocusing term, stabilization may be provided by the 2D
parabolic potential,
\begin{equation}
V\left( x,y\right) =\left( \Omega ^{2}/2\right) r^{2},  \label{Omega}
\end{equation}%
cf. Eq. (\ref{V}). In an analytical form, this possibility can be
demonstrated for small values of $\Omega ^{2}$ in Eq. (\ref{Omega}),
competing with small fractality parameter $\left( 2-\alpha \right) $, which
determines the proximity to the normal diffraction ($\alpha =2$).

First, the effect of the weak trapping potential can be taken into regard by
means of the respective VA for the NLSE including this potential, the
regular diffraction ($\alpha =2$), and cubic self-focusing \cite{Viskol}.
The VA for the stationary wave function with zero vorticity (see Eq. (\ref%
{Phi}) with $s=0$) is based on the 2D Gaussian ansatz,%
\begin{equation}
\mathcal{U}(r)=A\exp \left( -\frac{r^{2}}{2W^{2}}\right) ,  \label{ans2D}
\end{equation}%
with the norm (see Eq. (\ref{2D}))%
\begin{equation}
\mathcal{N}=\pi A^{2}W^{2},  \label{normVA}
\end{equation}%
cf. Eqs. (\ref{ans}) and (\ref{Nans}). The respective expression for the
effective Lagrangian is%
\begin{equation}
L_{\mathrm{eff}}^{\mathrm{(2D)}}=-\frac{\mu }{2}\mathcal{N}+\frac{\mathcal{N}%
}{4W^{2}}-\frac{\mathcal{N}^{2}}{8\pi W^{2}}+\frac{\Omega ^{2}}{4}\mathcal{N}%
W^{2},  \label{Leff2D}
\end{equation}%
cf. Eq. (\ref{Leff}). The Euler-Lagrange equations following from here, $%
\partial L_{\mathrm{eff}}^{\mathrm{(2D)}}/\partial \left( W,\mathcal{N}%
\right) =0$, lead to an expression showing how potential (\ref{Omega}) with
small $\Omega ^{2}$ lifts the norm degeneracy of the 2D Townes solitons:%
\begin{equation}
\mathcal{N}\approx 2\pi \left( 1-\frac{\Omega ^{2}}{4\mu ^{2}}\right) .
\label{lifting}
\end{equation}%
Here, the first term corresponds to the above-mentioned VA prediction for
the degenerate norm of the Townes solitons in the 2D\ cubic NLSE with the
normal diffraction ($\alpha =2$), see Eq. (\ref{VA-TS}). In the lowest
approximation, a small effect of $\Omega ^{2}$ may be neglected in the
corresponding VA-predicted relation between the soliton's width $W$ and
propagation constant $-\mu $ :%
\begin{equation}
W^{2}\approx -(2\mu )^{-1}.  \label{width}
\end{equation}

While Eq. (\ref{lifting}) pertains to $\alpha =2$, the expansion of the
free-space ($V=0$) scaling relation (\ref{N(mu)}) for small fractality, $%
0<2-\alpha \ll 1$, yields%
\begin{equation}
N\approx 2\pi \left[ 1-\frac{2-\alpha }{2}\ln \left( -\mu \right) \right] ,
\label{2-alpha}
\end{equation}%
where the value of $N$ for the Townes soliton at $\alpha =2$ is substituted
by the VA value (\ref{VA-TS}). Combining Eqs. (\ref{lifting}) and (\ref%
{2-alpha}) yields an expression which makes it possible to analyze the
competition of the stabilizing and destabilizing effects produced,
respectively, by the trapping potential and the weak fractality:
\begin{equation}
N\approx 2\pi \left[ 1-\frac{\Omega ^{2}}{4\mu ^{2}}-\frac{2-\alpha }{2}\ln
\left( -\mu \right) \right] .  \label{competing}
\end{equation}%
The stable part of the family of weakly-fractional solitons characterized by
dependence (\ref{competing}) may be identified by means of the VK criterion (%
\ref{VaKo}). It predicts that stable solitons are those with the propagation
constant $-\mu >0$ subject the following constraint:%
\begin{equation}
\mu ^{2}<\Omega ^{2}/\left( 2-\alpha \right) .  \label{final}
\end{equation}%
Thus, according to Eq. (\ref{width}), stable solitons should be wide
enough, $W^{2}>\sqrt{2-\alpha }/\left( 2\Omega \right) $. This conclusion is
natural, as the soliton should be sufficiently wide to feel the stabilizing
effect of the trapping potential.

\section{Conclusion}

Studies of the wave propagation in fractional media have made remarkable
progress, starting from the fractional linear Schr\"{o}dinger equation,
which was introduced by Laskin \cite{Lask1} for quantum-mechanical particles
moving by L\'{e}vy flights. The fractality in that equation is represented
by the derivative of the Riesz type, characterized by the L\'{e}vy index, $%
\alpha $. Actually, this is an integral operator (\ref{FracDefi}), defined
by means of the combination of direct and inverse Fourier transforms. This
operator may be approximately reduced to a combination of usual local
derivatives when it acts on the wave function built as a rapidly oscillating
continuous-wave carrier multiplied by a slowly varying envelope, as shown by
Eq. (\ref{quasi}).

The next step was the realization of the fractional Schr\"{o}dinger equation
as one governing the paraxial wave propagation in optical setups emulating
the fractional diffraction \cite{EXP3}. The implementation of the fractional
Schr\"{o}dinger equation in optics has made it natural to include the Kerr
(as well as non-Kerr) nonlinearities, thus arriving at the fractional NLSE.
The nonlinear equations render it possible to predict various self-trapped
modes, such as solitons and solitary vortices, in these settings. The
present article offers a brief review of some recent theoretical results on
this topic, while experimental observations of fractional solitons have not
been published, as yet. The modes considered in the review include basic 1D
solitons and 2D solitary vortices in the free space, supported by the cubic
and cubic-quintic nonlinearities, respectively. The 1D solitons are
considered in the interval of values of the L\'{e}vy index $1<\alpha \leq 2$
(the fractional NLSE with the cubic self-focusing term in 1D gives rise to
the collapse at $\alpha \leq 1$, while $\alpha =2$ corresponds to the usual
diffraction term). All 2D solitons supported by cubic nonlinearity in the
free space are unstable at $\alpha \leq 2$.

Also considered are 1D solitons in external potentials, which make it
possible to produce stable higher-order (multiple-peak) solitons and study
the spontaneous symmetry breaking in double-well potentials. In addition to
that, trapping potentials may stabilize 1D solitons at $\alpha <1$ and 2D
ones at $\alpha <2$. For the 2D setting, dynamics of necklace-shaped soliton
clusters carrying overall vorticity is included too. Proceeding to the
fractional cubic-quintic CGLE, the review briefly surveys 1D dissipative
solitons predicted by that equation.

This mini-review is far from being a comprehensive one. Among other topics
related to the fractional NLSEs are, \textit{inter alia}, $\mathcal{PT}$%
-symmetric solitons maintained by complex potentials subject to condition (%
\ref{PT}). In particular, breaking and restoration of the solitons' $%
\mathcal{PT}$ symmetry was recently considered in Ref. \cite{ghosts}. An
interesting possibility is to introduce fractional equations of the NLSE
type with quadratic or quadratic-cubic nonlinearity \cite{Thirouin,Liangwei}%
. Also theoretically considered were various modes supported by a
modulationally stable background field (continuous wave), such as dark
solitons and vortices, \textquotedblleft bubbles", and W-shaped solitons
\cite{we,we2}.

The theory may be developed in other directions. In particular, NLSEs of a
different type, that account for temporal propagation of optical waves under
the action of the fractional group-velocity dispersion, were developed in
Refs. \cite{Jorge1} and \cite{Jorge2}. The analysis has produced solitons in
those models as well.

\section*{Acknowledgments}

I highly appreciate collaborations with colleagues on various aspects of the
theory reviewed in this paper, especially with Jorge Fujioka, Shangling He,
Yingji He, Pengfei Li, Dumitru Mihalache, Jianhua Zeng, and Liangwei Zeng.
This work was supported, in part, by the Israel Science Foundation through
grant No. 1286/17.

\section*{Conflicts of Interest}

The author declares that there are no conflicts of interest related to this
article.


\begin{thebibliography}{99}
\bibitem{REV1} Y. S. Kivshar, B. A. Malomed, Dynamics of solitons in nearly
integrable systems, Rev. Mod. Phys. \textbf{61}(4), 763-915 (1989).

\bibitem{REV3} B. A. Malomed, D. Mihalache, F. Wise, L. Torner,
Spatiotemporal optical solitons. J. Opt. B \textbf{7}, R53-R72 (2005).

\bibitem{REV5} Y. V. Kartashov, B.A. Malomed, L. Torner, Solitons in
nonlinear lattices. Rev. Mod. Phys. \textbf{83}, 247-306 (2011).

\bibitem{REV6} Z. Chen, M. Segev, D. N. Christodoulides, Optical spatial
solitons: historical overview and recent advances. Rep. Prog. Phys. \textbf{%
75}, 086401 (2012).

\bibitem{REV8} B. A. Malomed, Multidimensional solitons: Well-established
results and novel findings. Eur. Phys. J. Spec. Top. \textbf{225}, 2507-2532
(2016).

\bibitem{REV9} Y. V. Kartashov, G. E. Astrakharchik, B. A. Malomed, L.
Torner, Frontiers in multidimensional self-trapping of nonlinear fields and
matter. Nat. Rev. Phys. \textbf{1}, 185-197 (2019).

\bibitem{REV10} B. A. Malomed, (INVITED)\ Vortex solitons: Old results and
new perspectives, Physica D \textbf{399}, 108-137 (2019).

\bibitem{REV11} D. Mihalache, Localized structures in optical and
matter-wave media: A selection of recent studies, Rom. Rep. Phys. \textbf{73}%
(2), 403 (2021).

\bibitem{Lask1} N. Laskin, Fractional quantum mechanics and L\'{e}vy path
integrals. Phys. Lett. A \textbf{268}, 298-305 (2000).

\bibitem{Lask2} N. Laskin, \textit{Fractional quantum mechanics} (World
Scientific: Singapore, 2018).

\bibitem{EXP1} B. A. Stickler, Potential condensed-matter realization of
space-fractional quantum mechanics: The one-dimensional L\'{e}vy crystal.
Phys. Rev. E \textbf{88}, 012120 (2013).

\bibitem{EXP2} F. Pinsker, W. Bao, Y. Zhang, H. Ohadi, A. Dreismann, and J.
J. Baumberg, Fractional quantum mechanics in polariton condensates with
velocity-dependent mass. Phys. Rev. B \textbf{92}, 195310 (2015).

\bibitem{EXP3} S. Longhi, Fractional Schr\"{o}dinger equation in optics.
Opt. Lett. \textbf{40}, 1117-1120 (2015).

\bibitem{PROP} Y. Zhang, X. Liu, M. R. Beli\'{c}, W. Zhong, Y. Zhang, M.
Xiao, Propagation dynamics of a light beam in a fractional Schr\"{o}dinger
equation, Phys. Rev. Lett. \textbf{115}(18), 180403 (2015).

\bibitem{PTS} Y. Zhang, H. Zhong, M. R. Beli\'{c}, Y. Zhu, W. Zhong, Y.
Zhang, D. N. Christodoulides, M. Xiao, $\mathcal{PT}$ symmetry in a
fractional Schr\"{o}dinger equation, Laser Photonics Rev. \textbf{10}(3),
526-531 (2016).

\bibitem{ghosts} P. Li, B. A. Malomed and D. Mihalache, Symmetry-breaking
bifurcations and ghost states in the fractional nonlinear Schr\"{o}dinger
equation with a $\mathcal{PT}$-symmetric potential, Opt. Lett. \textbf{46},
3267-3270 (2021).

\bibitem{Conti} L. Zhang, Z. He, C. Conti, Z. Wang, Y. Hu, D. Lei, Y. Li,
and D. Fan, Modulational instability in fractional nonlinear Schr\"{o}dinger
equation, Commun. Nonlin. Sci. Numer. Simulat. \textbf{48}, 531-540 (2017).

\bibitem{soliton1} S. Secchi and M. Squassina, Soliton dynamics for
fractional Schr\"{o}dinger equations, Applicable Analysis, \textbf{93},
1702-1729 (2014).

\bibitem{soliton2} S. Duo and Y. Zhang, Mass-conservative Fourier spectral
methods for solving the fractional nonlinear Schr\"{o}dinger equation,
Computers and Mathematics with Applications \textbf{71}, 2257-2271 (2016).

\bibitem{Frac1} W. P. Zhong, M. R. Beli\'{c}, B. A. Malomed, Y. Zhang, and
T. Huang, Spatiotemporal accessible solitons in fractional dimensions, Phys.
Rev. E \textbf{94}, 012216 (2016).

\bibitem{Frac2} W. P. Zhong, M. R. Beli\'{c}, and Y. Zhang, Accessible
solitons of fractional dimension, Ann. Phys. \textbf{368}, 110-116 (2016).

\bibitem{soliton3} Y. Hong and Y. Sire, A new class of traveling solitons
for cubic fractional nonlinear Schr\"{o}dinger equations, Nonlinearity
\textbf{30}, 1262-1286 (2017).

\bibitem{Chen} M. Chen, S. Zeng, D. Lu, W. Hu, and Q. Guo, Optical solitons,
self-focusing, and wave collapse in a space-fractional Schr\"{o}dinger
equation with a Kerr-type nonlinearity, Phys. Rev. E \textbf{98}, 022211
(2018).

\bibitem{Frac3} Q. Wang, J. Li, L. Zhang, and W. Xie, Hermite-Gaussian-like
soliton in the nonlocal nonlinear fractional Schr\"{o}dinger equation, EPL
\textbf{122}, 64001 (2018).

\bibitem{Frac4} Q. Wang, and Z. Z. Deng, Elliptic Solitons in
(1+2)-dimensional anisotropic nonlocal nonlinear fractional Schr\"{o}dinger
equation, IEEE Photonics J. \textbf{11}, 1-8 (2019).

\bibitem{Frac5a} C. Huang and L. Dong, Gap solitons in the nonlinear
fractional Schr\"{o}dinger equation with an optical lattice, Opt. Lett.
\textbf{41}, 5636-5639 (2016).

\bibitem{Frac5b} J. Xiao, Z. Tian, C. Huang, and L. Dong, Surface gap
solitons in a nonlinear fractional Schr\"{o}dinger equation, Opt. Express
\textbf{26}, 2650-2658 (2018).

\bibitem{soliton4} L. F. Zhang, X. Zhang, H. Z. Wu, C. X. Li, D. Pierangeli,
Y. X. Gao, and D. Y. Fan, Anomalous interaction of Airy beams in the
fractional nonlinear Schr\"{o}dinger equation, Opt. Exp. \textbf{27},
27936-27945 (2019).

\bibitem{Frac5c} L. Dong and Z. Tian, Truncated-Bloch-wave solitons in
nonlinear fractional periodic systems, Ann. Phys. \textbf{404}, 57-64 (2019).

\bibitem{Frac5} L. Zeng and J. Zeng, One-dimensional gap solitons in quintic
and cubic-quintic fractional nonlinear Schr\"{o}dinger equations with a
periodically modulated linear potential, Nonlinear Dyn. \textbf{98}, 985-995
(2019).

\bibitem{Frac6} P. Li, B. A. Malomed, and D. Mihalache, Vortex solitons in
fractional nonlinear Schr\"{o}dinger equation with the cubic-quintic
nonlinearity, Chaos Solitons Fract. \textbf{137}, 109783 (2020).

\bibitem{Frac7} Q. Wang and G. Liang, Vortex and cluster solitons in
nonlocal nonlinear fractional Schr\"{o}dinger equation, J. Optics \textbf{22}%
, 055501 (2020)

\bibitem{Frac8} L. Zeng and J. Zeng, One-dimensional solitons in fractional
Schr\"{o}dinger equation with a spatially periodical modulated nonlinearity:
nonlinear lattice, Opt. Lett. \textbf{44}, 2661-2664 (2019).

\bibitem{Frac9} Y. Qiu, B. A. Malomed, D. Mihalache, X. Zhu, X. Peng, and Y.
He, Stabilization of single- and multi-peak solitons in the fractional
nonlinear Schr\"{o}dinger equation with a trapping potential, Chaos Solitons
Fract. \textbf{140}, 110222 (2020).

\bibitem{Frac10} P. Li, B. A. Malomed, and D. Mihalache, Metastable soliton
necklaces supported by fractional diffraction and competing nonlinearities,
Opt. Exp. \textbf{28}, 34472-33488 (2020).

\bibitem{Frac11} L. Zeng, D. Mihalache, B. A. Malomed, X. Lu, Y. Cai, Q.
Zhu, and J. Li, Families of fundamental and multipole solitons in a
cubic-quintic nonlinear lattice in fractional dimension, Chaos Solitons
Fract. \textbf{144}, 110589 (2021).

\bibitem{Frac12} L. Zeng and J. Zeng, Preventing critical collapse of
higher-order solitons by tailoring unconventional optical diffraction and
nonlinearities, Commun. Phys. \textbf{3}, 26 (2020).

\bibitem{Frac13} M. I. Molina, The fractional discrete nonlinear Schr\"{o}%
dinger equation, Phys. Lett. A \textbf{384}, 126180 (2020)

\bibitem{Frac14} Y. Qiu, B. A. Malomed, D. Mihalache, X. Zhu, L. Zhang, and
Y. He, Soliton dynamics in a fractional complex Ginzburg-Landau model, Chaos
Solitons Fract. \textbf{131}, 109471 (2020).

\bibitem{Frac15} P. Li, J. Li, B. Han, H. Ma, and D. Mihalache, D.: $%
\mathcal{PT}$-symmetric optical modes and spontaneous symmetry breaking in
the space-fractional Schr\"{o}dinger equation, Rom. Rep. Phys. \textbf{71},
106 (2019).

\bibitem{Frac16} Li, P., Dai C.: Double Loops and Pitchfork Symmetry
Breaking Bifurcations of Optical Solitons in Nonlinear Fractional Schr\"{o}%
dinger Equation with Competing Cubic-Quintic Nonlinearities, Ann. Phys.
(Berlin) \textbf{532}, 2000048 (2020).

\bibitem{Frac17} P. Li, B. A. Malomed, and D. Mihalache, Symmetry breaking
of spatial Kerr solitons in fractional dimension, Chaos Solitons Fract.
\textbf{132}, 109602 (2020).

\bibitem{Frac18} L. Zeng, and J. Zeng, Fractional quantum couplers, Chaos
Solitons Fract, \textbf{140}, 110271 (2020).

\bibitem{Frac19} L. Zeng, J. Shi, X. Lu, Y. Cai, Q. Zhu, H. Chen, H. Long,
and J. Li, Stable and oscillating solitons of $\mathcal{PT}$-symmetric
couplers with gain and loss in fractional dimension. Nonlinear Dyn. \textbf{%
103}, 1831-1840 (2021).

\bibitem{Riesz} M. Cai and C. P. Li, On Riesz derivative, Fractional
Calculus and applied analysis \textbf{22}, 287-301 (2019).

\bibitem{PA} D. S. Petrov and G. E. Astrakharchik, Ultradilute
low-dimensional liquids, Phys. Rev. Lett. \textbf{117}, 100401 (2016).

\bibitem{Liangwei} L. Zeng, Y. Zhu, B. A. Malomed, D. Mihalache, Q. Wang, H.
Long, Y. Cai, X. Lu, and J. Li, Quadratic fractional solitons, to be
published.

%\bibitem{Yingji} Y. Qiu, B. A. Malomed, D. Mihalache, X. Zhu, L. Zhang, and
%Y. He, Soliton dynamics in a fractional complex Ginzburg-Landau model,
%Chaos, Solitons and Fractals \textbf{131}, 109471 (2020).

\bibitem{soliton5} P. Li, R. Li, and C. Dai, Existence, symmetry breaking
bifurcation and stability of two-dimensional optical solitons supported by
fractional diffraction, Opt. Exp. \textbf{29}, 3193-3210 (2021).

\bibitem{VK} N. G. Vakhitov and A. A. Kolokolov, Stationary solutions of the
wave equation in a medium with nonlinearity saturation. Radiophys. Quantum
Electron. \textbf{16}(7), 783-789 (1973).

\bibitem{Berge} L. Berg\'{e}, Wave collapse in physics: principles and
applications to light and plasma waves, Phys. Rep. \textbf{303}, 259-370
(1998).

\bibitem{we} L. Zeng, B. A. Malomed, D. Mihalache, Y. Cai, X. Lu, Q. Zhu,
and J. Li, Bubbles and W-shaped solitons in Kerr media with fractional
diffraction, Nonlinear Dynamics \textbf{104}, 4253-4264 (2021).

\bibitem{Baleanu} S. I. Muslih, O. P. Agrawal, and D. Baleanu, A Fractional
Schr\"{o}dinger equation and its solution, Int. J. Theor. Phys. \textbf{49},
1746-1752 (2010).

\bibitem{Anderson} M. Desaix, D. Anderson, and M. Lisak, Variational
approach to collapse of optical pulses, J. Opt. Soc. Am. B \textbf{8},
2082-2086 (1991).

\bibitem{Alexander} T. J. Alexander and L. Berg\'{e}, Ground states and
vortices of matter-wave condensates and optical guided waves, Phys Rev E
\textbf{65}, 026611 (2001).

\bibitem{PLA} D. Mihalache, D. Mazilu, B. A. Malomed and F. Lederer, Vortex
stability in nearly two-dimensional Bose-Einstein condensates with
attraction, Phys Rev. A \textbf{73}, 043615 (2006).

\bibitem{book} B. A. Malomed, editor: \textit{Spontaneous Symmetry Breaking,
Self-Trapping, and Josephson Oscillations}, (\noindent Springer-Verlag:
Berlin and Heidelberg, 2013).

\bibitem{Yingji1} S. He, B. A. Malomed, D. Mihalache, X. Peng, X, Yu, Y. He,
and D. Den, Propagation dynamics of abruptly autofocusing circular Airy
Gaussian vortex beams in the fractional Schr\"{o}dinger equation, Chaos,
Solitons \& Fractals \textbf{142}, 110470 (2021).

\bibitem{Yingji2} S. He, B. A. Malomed, D. Mihalache, X. Peng, Y. He, and D.
Deng, Propagation dynamics of radially polarized symmetric Airy beams in the
fractional Schr\"{o}dinger equation, Phys. Lett. A \textbf{404}, 127403
(2021).

\bibitem{Bulgaria} Kh. I. Pushkarov, D. I. Pushkarov, and I. V. Tomov,
Self-action of light beams in nonlinear media: soliton solutions, Opt.
Quant. Electr. \textbf{11}, 471-478 (1979).

\bibitem{necklace1} M. Solja\v{c}i\'{c} and M. Segev, Integer and fractional
angular momentum borne on self-trapped necklace-ring beams, Phys. Rev. Lett.
\textbf{86}, 420-423 (2001).

\bibitem{necklace2} A. S. Desyatnikov and Y. S. Kivshar, Necklace-ring
vector solitons, Phys. Rev. Lett. \textbf{87}, 033901 (2001).

\bibitem{necklace3} Y. V. Kartashov, L.-C. Crasovan, D. Mihalache, and L.
Torner, Robust propagation of two-color soliton clusters supported by
competing nonlinearities, Phys. Rev. Lett. \textbf{89}, 273902 (2002).

\bibitem{necklace-CQ} D. Mihalache, D. Mazilu, L.-C. Crasovan, B. A.
Malomed, F. Lederer, and L. Torner, Robust soliton clusters in media with
competing cubic and quintic nonlinearities, Phys. Rev. E \textbf{68}, 046612
(2003).

\bibitem{old} D. Mihalache, D. Mazilu, L.-C. Crasovan, I. Towers, A. V.
Buryak, B. A. Malomed, L. Torner, J. P. Torres, and F. Lederer, Stable
spinning optical solitons in three dimensions, Phys. Rev. Lett. \textbf{88},
073902 (2002).

\bibitem{Kruglov} V. I. Kruglov and R. A. Vlasov, Spiral self-trapping
propagation of optical beams, Phys. Lett. A \textbf{111}, 401-404 (1985).

\bibitem{Kruglov2} V. I. Kruglov, Yu. A. Logvin, and V. M. Volkov, The
theory of spiral laser beams in nonlinear media, J. Modern Opt. \textbf{39},
2277-2291 (1992).

\bibitem{Qin} J. Qin, G. Dong, and B. A. Malomed, Stable giant vortex annuli
in microwave-coupled atomic condensates, Phys. Rev. A \textbf{94}, 053611
(2016).

\bibitem{Viskol} Z. Chen, Y. Li, B. A. Malomed, and L. Salasnich,
Spontaneous symmetry breaking of fundamental states, vortices, and dipoles
in two and one-dimensional linearly coupled traps with cubic
self-attraction, Phys. Rev. A \textbf{96}, 033621 (2016).

\bibitem{Thirouin} J. Thirouin, On the growth of Sobolev norms of solutions
of the fractional defocusing NLS equation on the circle, Ann. Inst. H.
Poincare \textbf{AN34}, 509-531 (2017).

\bibitem{we2} L. Zeng, B. A. Malomed, D, Mihalache, Y. Cai, X. Lu, Q. Zhu,
and J. Li, Flat-floor bubbles, dark solitons, and vortices stabilized by
inhomogeneous nonlinear media, Nonlnear Dyn. \textbf{104}, 4253-4264 (2021).

\bibitem{Jorge1} J. Fujioka, A. Espinosa, and R. F. Rodr\'{\i}guez,
Fractional optical solitons, Phys. Lett. A \textbf{374}, 1126-1134 (2010).

\bibitem{Jorge2} J. Fujioka, A. Espinosa, R. F. Rodr\'{\i}guez, and B. A.
Malomed, Radiating subdispersive fractional optical solitons, Chaos \textbf{%
24}, 033121 (2014).
\end{thebibliography}
\end{document}